\documentclass[aps,prr,twocolumn,groupedaddress]{revtex4-2}

\usepackage{graphicx}
\usepackage{dcolumn}
\usepackage{bm}
\usepackage[dvipsnames]{xcolor}

\begin{document}

\def\gsim{\mathop {\vtop {\ialign {##\crcr 
$\hfil \displaystyle {>}\hfil $\crcr \noalign {\kern1pt \nointerlineskip } 
$\,\sim$ \crcr \noalign {\kern1pt}}}}\limits}
\def\lsim{\mathop {\vtop {\ialign {##\crcr 
$\hfil \displaystyle {<}\hfil $\crcr \noalign {\kern1pt \nointerlineskip } 
$\,\,\sim$ \crcr \noalign {\kern1pt}}}}\limits}


\title{Monte Carlo study on critical exponents of the classical Heisenberg model in ferromagnetic icosahedral quasicrystal}


\author{Shinji Watanabe$^1$, Tsunetomo Yamada$^2$, Hiroyuki Takakura$^3$, and Nobuhisa Fujita$^4$}
\affiliation{$^1$Department of Basic Science, Kyushu Institute of Technology, Kitakyushu, 804-8550, Fukuoka, Japan
\\
$^2$Department of Applied Physics, Tokyo University of Science, Niijuku, 125-8585, Tokyo, Japan
\\
$^3$Division of Applied Physics, Faculty of Engineering, Hokkaido University, Sapporo, 060-8628, Hokkaido, Japan
\\
$^4$Institute of Multidisciplinary Research for Advanced Materials, Tohoku University, Sendai, 980-8577, Miyagi, Japan
}


\date{\today}

\begin{abstract}
Quasicrystals (QCs) lack three-dimensional periodicity of atomic arrangement but possess long-range structural order, which are distinct from periodic crystals and random systems. 
Here, we show how the ferromagnetic (FM) order arises in the icosahedral QC (i-QC) on the basis of the Monte Carlo simulation of the Heisenberg model on the Yb lattice of Cd$_{5.7}$Yb composed of regular icosahedrons. 
By finite-size scaling of the Monte Carlo data, we identified the critical exponents of the magnetization, magnetic susceptibility, and spin correlation length, $\beta=0.508(30)$, $\gamma=1.361(59)$, and $\nu=0.792(17)$, respectively.  We confirmed that our data satisfy the hyperscaling relation and estimated the other critical exponents $\alpha=-0.376(51)$, $\delta=3.68(23)$, and $\eta=0.282(65)$. 
These results show a new universality class inherent in the i-QC, which is different from those in periodic magnets and spin glasses. In the i-QC, each Yb site at vertices of the regular icosahedrons is classified into 8 classes with respect to the coordination numbers of the nearest-neighbor and next-nearest-neighbor bonds. We revealed the FM-transition mechanism by showing that the difference in the local environment of each site is governed by cooperative evolution of spin correlations upon cooling, giving rise to the critical phenomena. 
\end{abstract}


\maketitle

\section{Introduction}

Magnetism has been one of the central subject in the condensed matter physics. Critical exponents are distinguishing parameters characterizing continuous phase transitions. A great deal of effort has been devoted to critical phenomena of magnetic transitions over years and the critical exponents have been identified in periodic magnets, which has established the universality class~\cite{Fisher,Kadanoff,Stanley,CL_text}. On the other hand, in disordered systems, the spin alignment is frozen at low temperatures, where spin glasses are formed~\cite{BY,Mydosh,KT_text}. The critical exponents at the spin-freezing temperature in disordered magnets were also determined and the universality class of the spin glass was identified~\cite{Viet2009,Ogawa}. 

Quasicrystals (QCs) are a new class of solids, which have no three-dimensional periodicity of atoms but possess the long-range structural order~\cite{Levine1984}. Hence, crystallographically QC is distinct from periodic crystal and random system.  
After the discovery of QC in 1984~\cite{Shechtman}, a stable binary QC was first synthesized, that is the single crystal of Cd$_{5.7}$Yb~\cite{Tsai}. 
The structure model of atoms was solved by one of the present authors~\cite{Takakura} and it was revealed that Cd$_{5.7}$Yb is composed of regular icosahedrons and such QCs are referred to as icosahedral QCs (i-QCs). 
Then, binary compounds of rare-earth based i-QCs Cd-R (R=Gd, Tb, Dy, Ho, Er and Tm)~\cite{Goldman} and ternary compounds Au-SM-R (SM=Ga, Al, R=Yb, Gd, Tb, and Dy) were synthesized~\cite{Deguchi,Tamura2021,Takeuchi2023}. 

Notable is that the i-QCs have played a significant role in advancing condensed matter physics~\cite{Tsai1990,Takeuchi1993,Sato2022}.
These developments of materials led to discoveries of quantum critical phenomena in the i-QC Au-Al-Yb~\cite{Deguchi} as well as superconductivity in the i-QC Al-Zn-Mg~\cite{Kamiya} and lattice dynamics in the i-QC Al-Pd-Mn~\cite{Matsuura} and i-QC Zn-Mg-Sc~\cite{Marc}, which have brought about paradigm shift of fundamental concept of physics ever established in periodic crystals. 

One of the long-standing issues unresolved for QCs has been whether magnetic long-range order is realized in QCs. Despite intensive studies for nearly four decades, no magnetic long-range order such as ferromagnetism but only spin-glass behavior has been observed in rare-earth-based i-QCs, e.g., Cd-R (R=Gd, Tb, Dy, Ho, Er and Tm)~\cite{Goldman}. Recent discovery of the FM long-range order in the i-QCs Au-Ga-R (R=Gd, Tb, and Dy)~\cite{Tamura2021,Takeuchi2023} and the antiferromagnetic order in the i-QC Au-In-Eu~\cite{Tamura2025} calls for theoretical investigation of the ordering mechanism and the property.
The experimental observations of the magnetization and the magnetic susceptibility in the FM phase of the i-QC Au-Ga-Dy have reported the critical exponents $\beta=0.54$ and $\gamma=0.89$~\cite{Takeuchi2023}.

The magnetism in the approximant crystals (ACs), which are the periodic crystals with the common local atomic configurations to those in the QC, has been extensively studied~\cite{Suzuki}.  
Recently, experimental identification of the critical exponents has been reported in the Gd-based 1/1 ACs~\cite{Shiino2022}. 
In the AC Au$_{72.7}$Si$_{13.6}$Gd$_{13.7}$,  $\beta=0.47$, $\gamma=1.12$, and $\delta=3.60$ were identified and in the AC Au$_{68.6}$Si$_{16.0}$Gd$_{15.4}$,  $\beta=0.51$, $\gamma=1.00$, and $\delta=3.38$ were identified~\cite{Shiino2022}. These observations in the i-QC and ACs indicate that the critical exponents of the magnetization $\beta$ are close to the mean-field value $\beta=0.5$, which also calls for theoretical study. 

So far, most of the theoretical studies on the magnetism in QCs have been devoted to the low-spatial dimensions~
\cite{Achiam,Tsunetsugu1987,Stinchcombe,Jeon2025,Godreche,Bhattacharjee,Okabe,Sorensen,Ledue1995,Ledue1997,Wessel2003,Vedmedenko2003,Vedmedenko2004,Thiem2015,Thiem,Jan2007,Komura,Koga2017,Koga2021,Koga2022,Okabe2024}.  
The Ising model on the two-dimensional (2D) QCs such as Penrose lattice was extensively studied and it was reported that the critical exponents are the same as those in the periodic square lattice~\cite{Bhattacharjee,Ledue1997,Thiem,Jan2007,Komura,Okabe2024}. 
In one-dimensional and 2D QCs, the properties of the magnetic excitations were investigated~\cite{Ashraff1989,Ashraff1990,Wessel2005,Inoue2020,Yamamoto2024}. 
Until a few years ago, theoretical study of the magnetism in three-dimensional (3D) QCs was not reported except for numerical studies on small clusters~\cite{Axe2001,Kons2005,Hucht2011,STS} and symmetry analysis of the magnetic i-QC~\cite{Lifshitz1998,Lifshitz2000}.

Recently, the formulation of the crystalline electric field (CEF) in the rare-earth-based QC has been succeeded on the basis of the point charge model~\cite{WK2021}. This has made it possible to clarify the magnetic anisotropy arising from the CEF for each rare-earth based i-QC and AC~\cite{WPNAS,WI2023}. The noncollinear Ising model on the i-QC taking into account the magnetic easy axis due to the CEF has theoretically been studied~\cite{WPNAS,WSR}. Noncollinear magnetic state and topological magnets characterized by topological number such as the hedgehog state and whirling-moment state have been theoretically shown, which suggests a possibility of i-QC and AC as potential magnetic materials. 
Magnetic excitations and their dynamics in the non-collinearly ordered state in the i-QC have also been revealed theoretically~\cite{WSR2022,WSR2022HH,WSR2023}. 

In the Gd-based i-QC and Eu-based i-QC, where Gd$^{3+}$ and Eu$^{2+}$ with $4f^7$ configuration have the ground multiplet $^{8}S_{7/2}$, the magnetic anisotropy arising from the CEF is expected to be absent since the total orbital angular momentum is zero, $L=0$. 
In this study, to gain insight into the magnetism in the Gd- and Eu-based i-QCs, we perform Monte Carlo simulation for the classical Heisenberg model on the i-QC. By analyzing the finite-size scaling of the Monte Carlo data,, we identify the critical exponents.

Organization of this paper is as follows. In Sec. II, we explain the Yb-lattice structure of the i-QC Cd$_{5.7}$Yb and introduce the spin model. In Sec. III, we explain the Monte Carlo method implemented in the present study. In Sec. IV, the results of the Monte Carlo simulation are presented. By the finite-size scaling of the Monte Carlo data, we identify the critical exponents. We analyze how the FM occurs in the i-QC by showing evolution of the magnetization as well as the spin correlation at each site under the different environment as temperature decreases. In Sec. V, we summarize our results and discuss the relevance to experiments.

\section{Model on icosahedral quasicrystal}



\begin{figure}[tb]
\centering
\includegraphics[width=7cm]{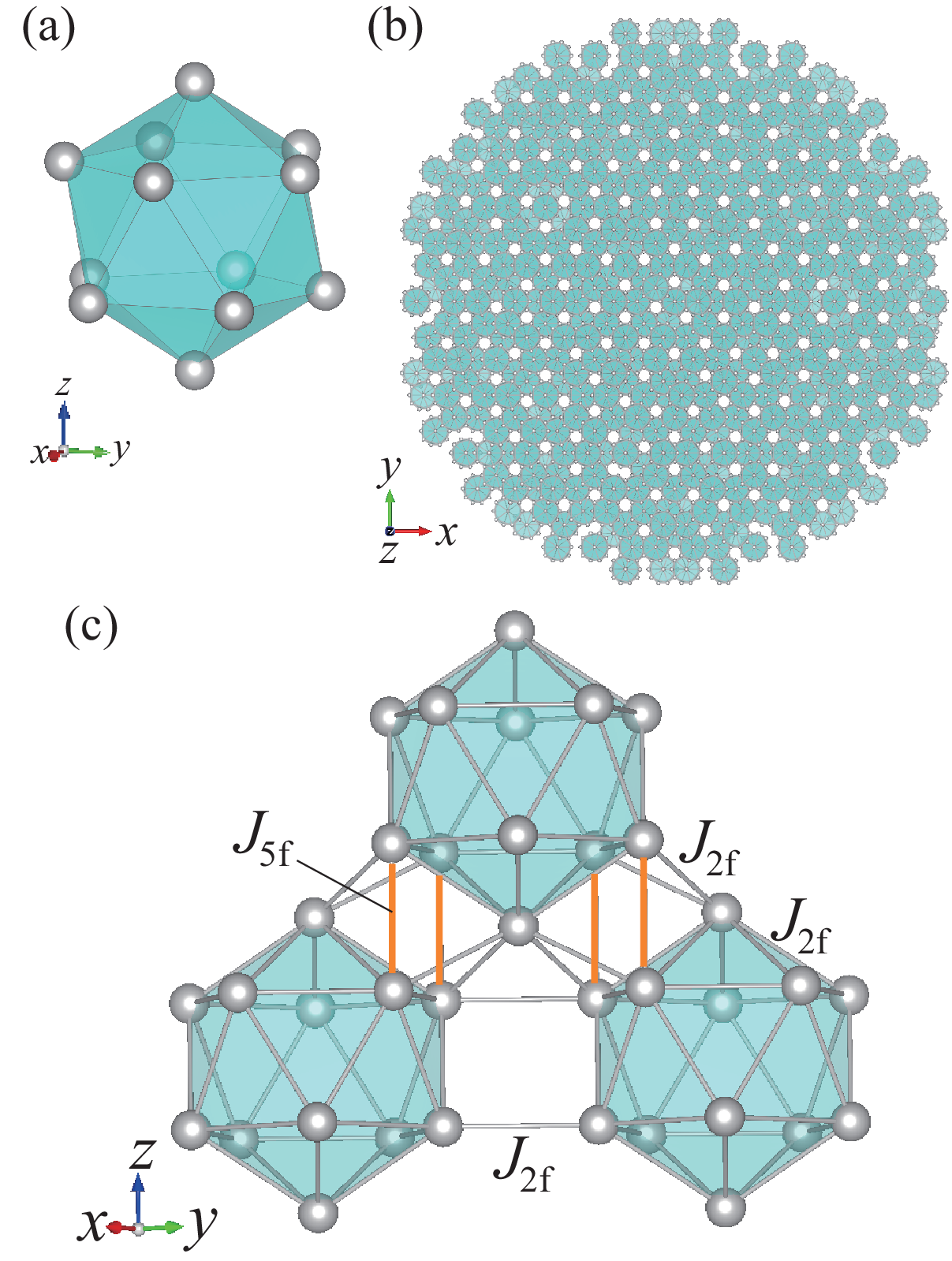}
\caption{(a) The regular icosahedron at 12 vertices of which Yb atom is located.  
(b) The Yb atoms in the icosahedral QC Cd$_{5.7}$Yb for $N=20364$ viewed from the $z$ axis direction.  
(c) Regular icosahedrons, at 12 vertices of which the Yb atoms are located, in the QC Cd$_{5.7}$Yb. The N.N. bond (orange) with the length being $a_{\rm ico}$ is along the 5-fold axis direction. The N.N.N. bond (gray) with the length being $5.9821~{\rm \AA}$ is along the 2-fold axis direction. The $z$ axis is set along the 5-fold axis direction. 
}
\label{fig:IC_QC}
\end{figure}

Let us start with the lattice structure of the i-QC Cd$_{5.7}$Yb~\cite{Takakura}. 
On the basis of the method described in Ref.~\cite{Takakura}, we generated the Yb sites of the i-QC Cd$_{5.7}$Yb with the icosahedral lattice parameter $a_{\rm ico}=5.6893~{\rm \AA}$. 
The Yb sites are located at 12 vertices of the regular icosahedron [see Fig.~\ref{fig:IC_QC}(a)] and also exists as a pair inside the acute rhombohedron. The ratio of the former Yb site and latter are about $70~\%$ and $30~\%$ respectively. 
In the present study, we ignored the latter sites as a first step of analysis, since the icosahedron is a common motif in both i-QC and ACs, dominating their magnetism~\cite{Suzuki}. 
Namely, we consider all the Yb atoms located at the vertices of the regular icosahedrons in the i-QC Cd$_{5.7}$Yb. 

In Fig.~\ref{fig:IC_QC}(a), the bonds connecting the neighboring vertices are along the 2-fold axis with the edge length of the icosahedron being $5.9821~{\rm \AA}$. The diagonal direction penetrating the two opposite vertices, e.g., the $z$ axis direction, is along the 5-fold axis. 
In Fig.~\ref{fig:IC_QC}(b), we show the lattice structure of Yb atoms in the i-QC Cd$_{5.7}$Yb for $N=20364$ with $N$ being the number of Yb sites, which is viewed from the $z$ axis, i.e., the 5-fold axis. 

Here we explain the way of setting the coordinate, as follows. 
The basic vector of the i-QC is directed to each vertex from the center of the icosahedron, which is given by  
\begin{eqnarray}
{\bm a}_1&=&(0,0,a_{\rm ico}), 
\nonumber
\\
{\bm a}_j&=&\left(a_{\rm ico}\sin{\theta}\cos\left(\frac{2\pi(j-2)}{5}\right), a_{\rm ico}\sin{\theta}\sin\left(\frac{2\pi(j-2)}{5}\right), \right.
\nonumber
\\ & & 
\left.
a_{\rm ico}\cos{\theta}\right),   (j=2,...,6),
\nonumber 
\end{eqnarray}
where $\theta$ is given by $\theta = \arccos(1/\sqrt{5})$. In the present study, we set the $z$ axis along the ${\bm a}_1$ direction. Then, the $x$ axis is taken along the vector obtained by setting the $z$ component of ${\bm a}_2$ zero, i.e., $(a_{\rm ico}\sin{\theta}, 0, 0)$. The $y$ axis is set to be perpendicular to the $z$ and $x$ axes. Namely, the view from the $z$ axis direction is along the 5-fold axis direction [see Fig.~\ref{fig:IC_QC}(b)], as noted above. 

We consider the Heisenberg model
\begin{eqnarray}
H=-\sum_{\langle i,j\rangle}J_{ij}{\bm S}_i\cdot{\bm S}_j,  
\label{eq:H}
\end{eqnarray}
where the spin is located at the Yb site and $J_{ij}$ is taken as $J_{\rm 5f}$ for the nearest-neighbor (N.N.) sites with the bond length $5.6893~{\rm \AA}$ and $J_{\rm 2f}$ for the next nearest-neighbor (N.N.N.) sites with the bond length $5.9821~{\rm \AA}$ [see Fig.~\ref{fig:IC_QC}(c)]. The former and the latter bonds are along the 5-fold axis and 2-fold axis directions, respectively. The spin ${\bm S}_i=(S_{ix},S_{iy},S_{iz})$ at the $i$th site is treated as classical spin with $|{\bm S}_i|=1$. 
We study the FM interaction case with $J=J_{\rm 5f}=J_{\rm 2f}=1$. 



In QC, local environment is different site by site in general.  
In the i-QC Cd$_{5.7}$Yb, each Yb site located at vertices of the regular icosahedrons can be classified into 8 classes by the coordination numbers with respect to the N.N. bonds and the N.N.N. bonds~\cite{Kumazawa}. 

\begin{figure}[h]
\centering
\includegraphics[width=8cm]{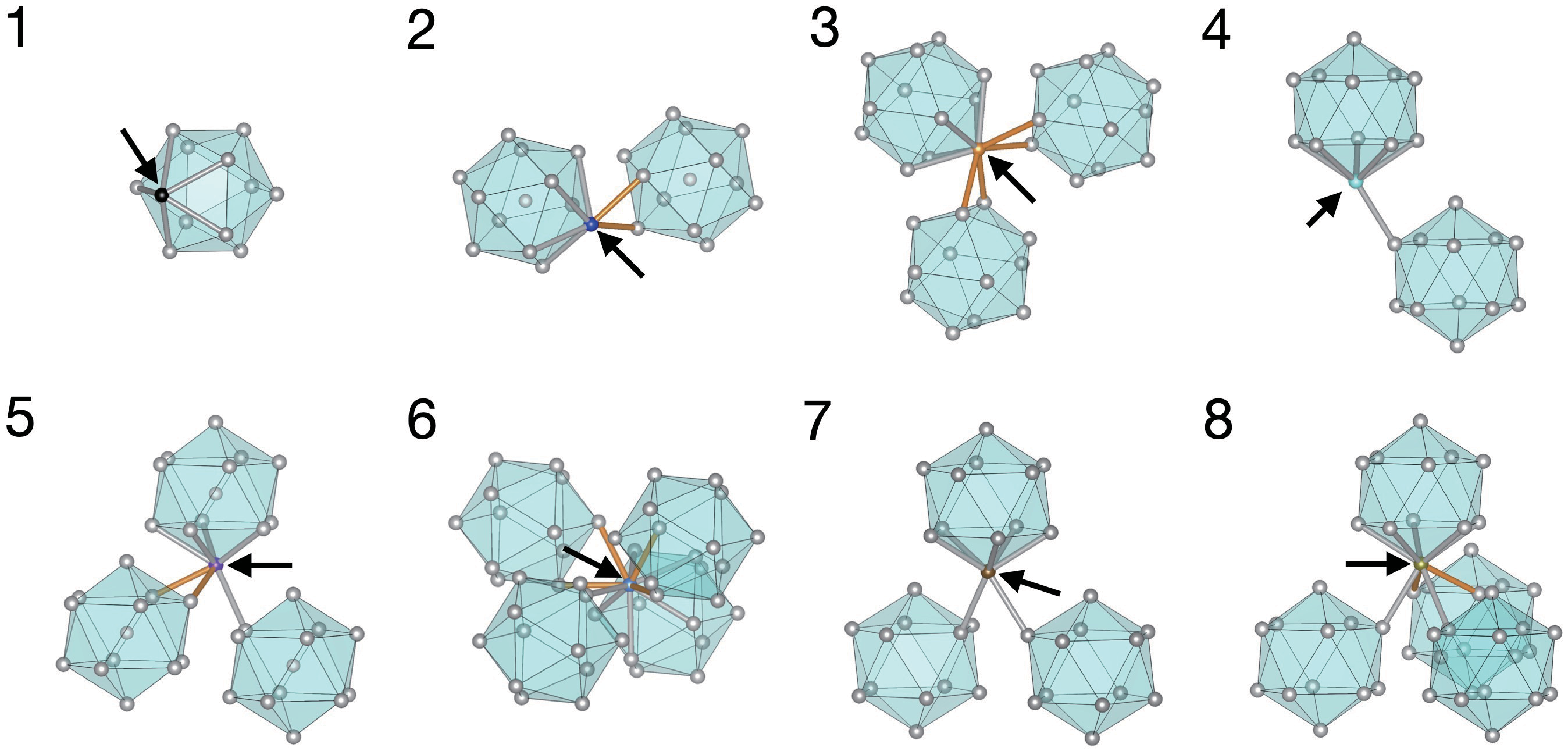}
\caption{
Local environment of 8 site class $(\lambda=1, 2, \cdots, 8)$. The number indicates the site class $\lambda$. The site indicated by arrow is the site which belongs to the class $\lambda$.  The orange bond denotes the N.N. bond along the 5-fold axis direction and the gray bond denotes the N.N.N. bond along the 2-fold axis direction. 
}
\label{fig:LE}
\end{figure}

Figure~\ref{fig:LE} shows the 8 local configurations for each Yb site, which was determined by using the hyperspace formalism, in the lattice structure consisting of the Yb icosahedral shells~\cite{Kumazawa}. Each configuration is characterized by the number of the N.N. and N.N.N. Yb sites, as shown in Table~\ref{tb:LE}. The frequency of each is also given in the table.


\begin{table}[h]
\begin{tabular}{cccc}
\hline
class & number of N.N. site & number of N.N.N. site & frequency $(\%)$\\
\hline
1 & 0 & 5 & 3.000805 \\
2 & 2 & 5 & 9.3055392 \\
3 & 4 & 5 & 23.468722 \\
4 & 0 & 6 & 0.718494 \\
5 & 2 & 6 & 1.4459422 \\
6 & 4 & 6 & 28.880961 \\
7 & 0 & 7 & 10.433782 \\
8 & 2 & 7 & 22.745751 \\
\hline
\end{tabular}
\caption{ Local configuration of Yb atoms. The serial numbers in the first column represent each local configuration. The numbers of the N.N and N.N.N Yb site for each configuration are listed at the second and third column, respectively. The frequency of each configuration is listed at the fourth column (in percentage).}
\label{tb:LE}%
\end{table}

\begin{figure}[bt]
\centering
\includegraphics[width=7cm]{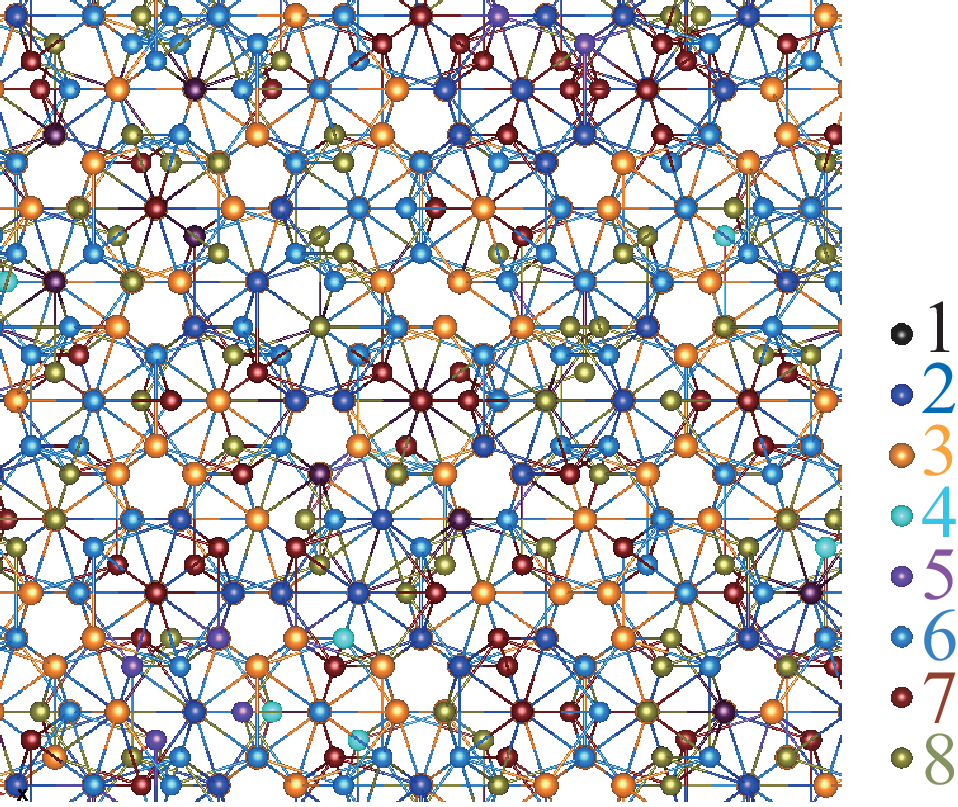}
\caption{ 
Each site is classified into 8 classes which are represented by 8 colors. 
The view is from the $z$ axis (5-fold axis) direction. 
}
\label{fig:QC_8_class}
\end{figure}

We show 8-classified sites illustrated by 8 different colors in Fig.~\ref{fig:QC_8_class}, which is the enlargement of the central part of Fig.~\ref{fig:IC_QC}(b), i.e., the $N=20364$ system viewed from the 5-fold axis direction. 
It looks that different site class is distributed in a intermixed manner. This is in sharp contrast to periodic crystals where each site classified as different site class is distributed periodically if ever.  On the other hand, in the i-QC, the site distribution is not completely random, but obeys a regular rule, which is the quasi periodicity with self similarity.  
In the random magnet with competing interactions, it is known that spin distribution is frozen at low temperatures, giving rise to the spin glass.  

In the present system, the collinear FM state is expected to be realized in the ground state because the FM interactions for the N.N. bonds and the N.N.N. bonds can contribute to cause the collinear alignment of all spins without any frustration. On the other hand, at high temperature, the spins are expected to be directed randomly by thermal fluctuations, which is expected to give rise to the paramagnetic phase. These speculations lead to the following interesting questions: How the present system composed of different site class shown in Fig.~\ref{fig:QC_8_class} undergo to the FM order as temperature decreases? Is there any difference in the nature of the phase transition from the periodic magnet and the spin glass? To address these issues, we study the Heisenberg model (\ref{eq:H}) on the i-QC by the Monte Carlo method.

\section{Monte Carlo method}
\label{sec:MC}

We perform the Monte Carlo simulation based on the heat-bath method~\cite{Miyatake} and the over-relaxation method~\cite{Creutz,Alonso}, which are combined with the replica-exchange method of parallel tempering~\cite{Hukushima}. 
We set 48 replicas which have different temperatures. In each replica, we perform the heat-bath single spin flip update followed by four times overrelaxation steps for all sites.
The combination of one heat-bath sweep and 4 over-relaxation sweeps constitutes our unit Monte Carlo step (mcs). 

In our Monte Carlo simulations, starting from the spin configuration taken to be random, we performed the $10^5$ mcs, 
during which the spin state and the energy at the temperature of neighboring replicas are exchanged each other following the replica exchange algorithm~\cite{Hukushima}. 
After this thermalization process, we perform the $10^5$ mcs 
for all spins with the replica exchange for parallel tempering,  
where the physical quantities are observed. We performed this 5 times independently by inputting different random-number seed for the initial spin configuration and evaluated the average value of the physical quantities with error bars.

We checked the mcs dependence of the physical quantities and confirmed that the mcs $10^5$ is sufficient (see Appendix~\ref{sec:MC_check}).

The correlation length of spins at the sites belonging to the same site class $\lambda$ is defined by~\cite{Cooper,Viet2009}
\begin{eqnarray}
\xi_{\lambda}=\frac{1}{2\sin\left(\frac{k_m}{2}\right)}\sqrt{\frac{M_{\lambda}({\bm 0})^2}{M_{\lambda}({\bm k})^2}-1}, 
\label{eq:corrl}
\end{eqnarray}
where ${\bm k}=(0,0,k_m)$ with $k_m=2\pi/L$ and $\lambda$ labels the site class $(\lambda=1,2,\cdots, 8)$.
Here we calculate the spin correlation length along the 5-fold axis which is taken as the $z$ axis and $L$ is the diameter of the system (see Fig.~S1c). 
In Eq.~(\ref{eq:corrl}), $M_{\lambda}({\bm k})^2$ is given by 
\begin{eqnarray}
M_{\lambda}({\bm k})^2=\sum_{\mu=x,y,z}\left|\frac{1}{N}\sum_{i=1}^{N_{\lambda}}S_{i\mu}e^{i{\bm k}\cdot{\bm r}_i}\right|^2,
\label{eq:def_m2}
\end{eqnarray}
where the summation $\sum_i$is taken over the sites belonging to the same site class $\lambda$. The total number of sites is the summation of each number of the site class $N=\sum_{\lambda=1}^{8}N_{\lambda}$.

\section{Results of Monte Carlo simulation}

We performed the Monte Carlo simulation for the $N=600$, 3168, 10440, 20364, 26412, 30048, 39360, 47520, and 62868 systems. In this section, we present the results and discuss the nature of the FM transition in the i-QC. In Sec.~\ref{sec:Binder}, we discuss the identification of the transition temperature $T_{\rm c}$ to the FM order in the bulk limit. In Sec.~\ref{sec:inner}, we explain the way to reduce the surface effect in the Monte Carlo sampling. In Sec.~\ref{sec:ECMchi}, the temperature dependences of the internal energy, specific heat, magnetization, and magnetic susceptibility are presented. Then, the critical exponents of spin correlation length $\nu$, the magnetization $\beta$, the magnetic susceptibility $\gamma$, and the specific heat $\alpha$ are identified in Sec.~\ref{sec:spin_corr}, Sec.~\ref{sec:beta}, Sec.~\ref{sec:gamma}, and Sec.~\ref{sec:alpha}, respectively. In Sec.~\ref{sec:UC}, we discuss the universality class of the Heisenberg model on the i-QC and make a comparison with those on the periodic crystals and the spin glass. In Sec.~\ref{sec:M_LE}, we analyze the magnetization and spin correlation length for each site classified by the coordination number, which is characteristic of the i-QC. 

\subsection{Binder parameter and transition temperature}
\label{sec:Binder}

To identify the magnetic transition temperature $T_{\rm c}$ in the bulk limit, the Binder parameter~\cite{Binder}  
\begin{eqnarray}
U=1-\frac{1}{3}\frac{\langle M^4\rangle}{\langle M^2\rangle}
\label{eq:BR}
\end{eqnarray}
is known to be useful~\cite{Binder}, where $M$ is the magnetization defined by 
\begin{eqnarray}
M=\sqrt{M_x^2+M_y^2+M_z^2}
\label{eq:M}
\end{eqnarray} 
with 
\begin{eqnarray}
M_{\mu}=\frac{1}{N}\sum_{i=1}^{N}S_{i\mu}  \ (\mu=x, y, z).
\label{eq:M_mu}
\end{eqnarray}
In the high-temperature limit, $U$ can be calculated by considering Gaussian fluctuations around ${\bm M}={\bf 0}$
as $U\to \frac{4}{9}$. 
For low temperatures, the magnetic order occurs. 
In the collinear FM phase, $U$ is shown to be $U\to \frac{2}{3}$ for $T\to 0$.  

The Binder parameter has the scaling form as~\cite{Binder} 
\begin{eqnarray}
U(T,L)=g(L^{1/\nu}(T-T_{\rm c})),  
\label{eq:BR_g}
\end{eqnarray}
where $L$ is the linear size of the system and $\nu$ is the critical exponent for the spin-spin correlation length. 

\begin{figure}[bt]
\centering
\includegraphics[width=7cm]{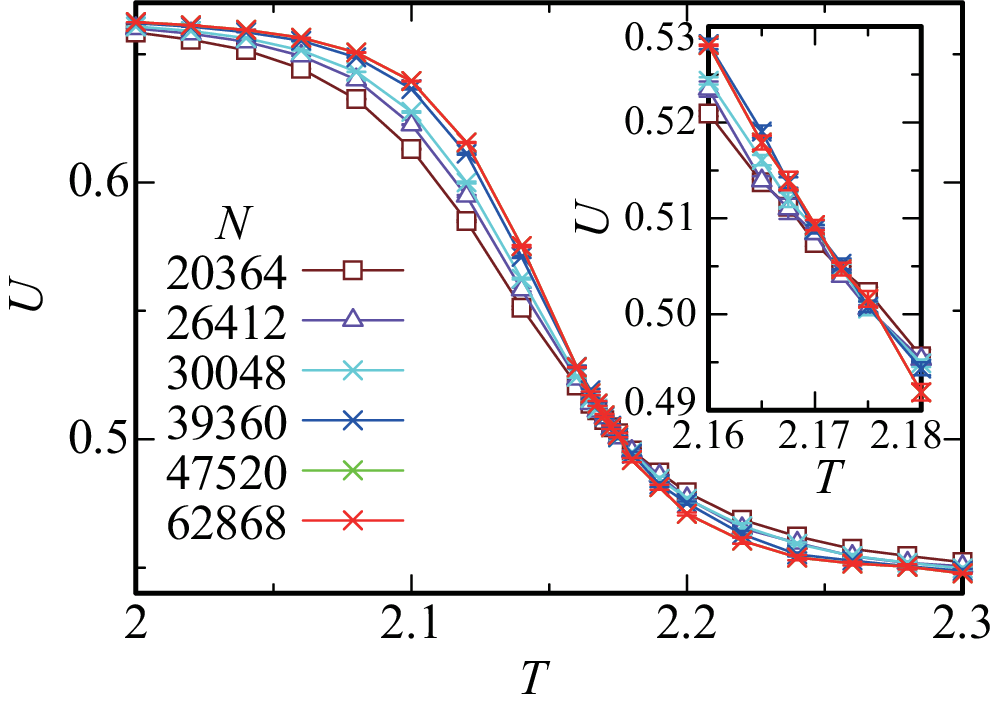}
\caption{
Temperature dependence of Binder parameter $U$ calculated for $J_{\rm 5f}=1$ and $J_{\rm 2f}=1$ in the $N=20364$, 26412, 30048, 39360, 47520 and 62868 systems. 
Inset is enlargement in the vicinity of $T=2.17$. 
}
\label{fig:U_T}
\end{figure}

We calculated the Binder parameter $U(T,L)$ for several system sizes. 
The results are shown in Fig.~\ref{fig:U_T}. We see that the crossing of $U(T,L)$s occur around $T\approx 2.17$.  
According to Eq.~(\ref{eq:BR_g}),  
at just the transition temperature, i.e., $T=T_{\rm c}$, the system size dependence vanishes in $U(T_{\rm c},L)$.  
This implies that the temperature at the crossing point of $U(T,L)$ for different system sizes gives the transition temperature $T_{\rm c}$.
The inset of Fig.~\ref{fig:U_T} is the enlargement of $U(T,L)$ in the vicinity of $T=2.17$. 
We see that the data for large system sizes with $N=20364$, 26412, 30048, 39360, 47520, and 62868 cross within error bars at $T=2.1725$, which indicates 
\begin{eqnarray}
T_{\rm c}=2.1725.
\label{eq:Tc}
\end{eqnarray}

Next, we proceed to analyze the critical behavior near $T=T_{\rm c}$. 
In this study, we focus on the bulk properties. Hence, before going into the detailed analysis, in the next subsection, we discuss the way to reduce the surface effect in the Monte Carlo simulation.

\subsection{Inner sites except for surface icosahedrons}
\label{sec:inner}

\begin{figure}[bt]
\centering
\includegraphics[width=7cm]{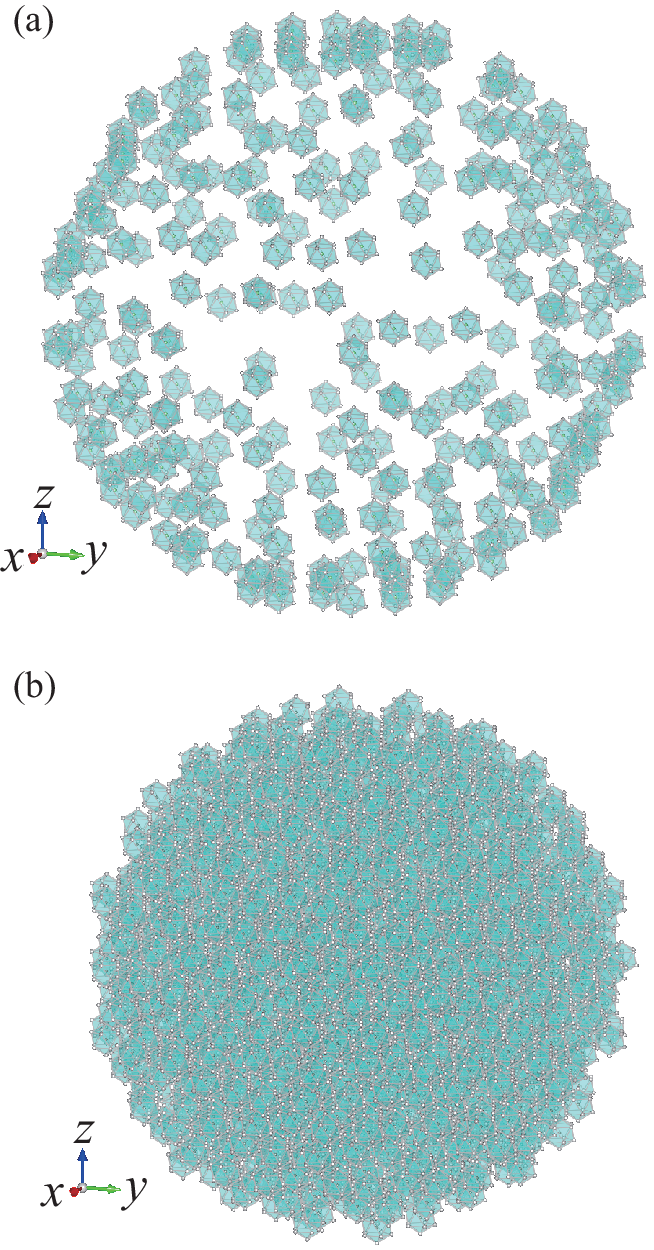}
\caption{
The Yb atoms in the icosahedral QC Cd$_{5.7}$Yb for 
(a) the outer sites and (b) the inner sites for $N=20364$. 
}
\label{fig:IN_OUT}
\end{figure}

To reduce the effect from the surface,
we calculate the average of physical quantities for the inner sites of the cluster in the Monte Carlo simulation, as recently done in the two-dimensional aperiodic lattice~\cite{Okabe}. Namely, we separate the two parts of the Yb atoms in the QC Cd$_{5.7}$Yb. One is located at the vertices of the outer icosahedrons (Fig.~\ref{fig:IN_OUT}a) and the other is those located at the vertices of the inner icosahedrons (Fig.~\ref{fig:IN_OUT}b). For example, in the $N=20364$ system, the numbers of the outer and inner icosahedrons are 284 and 1413 respectively where the numbers of the Yb sites are $N_{\rm OUT}=3408$ and $N_{\rm IN}=16956$ respectively. 

We performed the Monte Carlo simulation for $J_{\rm 5f}=1$ and $J_{\rm 2f}=1$ in the model (1) on the cluster with $N$ sites under the open boundary condition by updating the spins on all the sites and take the average of physical quantities for the inner $N_{\rm IN}$ sites. 
We compared the results of the averaged values of the magnetization, magnetic susceptibility, and specific heat for the inner $N_{\rm IN}$ sites and the total $N$ sites. 
The results for $N=62868$ are presented and also the results for $N=20364$ are noted in Appendix~\ref{sec:SF}. 
The data calculated for the total sites and the inner sites show the similar temperature dependences but it turned out that there exist a slight deviation (see Fig.~\ref{fig:SF_comp} in Appendix~\ref{sec:SF}). 

Therefore, in the following subsections, we will discuss the average of the physical quantities calculated for the inner sites to reduce the surface effects. 
We list the numbers of the total sites $N$ and the inner sites $N_{\rm IN}$ which we will use in this Monte Carlo simulation in Table~\ref{tb:SF}. 

\begin{table}[b]
\begin{tabular}{cc}
\hline
Total $N$&  \ \ \  Inner $N_{\rm IN}$ \\
\hline
20364 & 16956 \\
26412 & 21864 \\
30048 & 26412  \\
39360 & 25760  \\
47520 & 39360  \\
62868 & 55548  \\
\hline
\end{tabular}
\caption{Number of total sites $N$ and innner sites $N_{\rm IN}$}
\label{tb:SF}%
\end{table}


\subsection{Temperature dependences of internal energy, specific heat, magnetization, and magnetic susceptibility}
\label{sec:ECMchi}

\begin{figure}[bt]
\centering
\includegraphics[width=7cm]{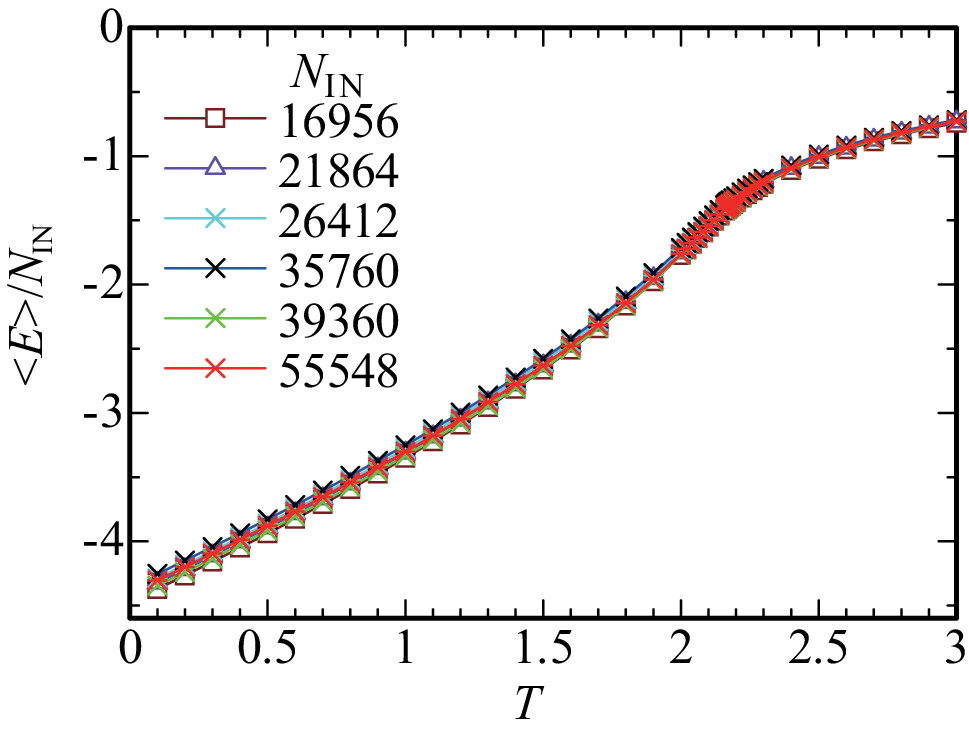}
\caption{
Temperature dependence of internal energy per site calculated for $J_{\rm 5f}=1$ and $J_{\rm 2f}=1$ in the $N=20364$, 26412, 30048, 39360, 47520 and 62868 systems.
}
\label{fig:E_T}
\end{figure}

\begin{figure}[bt]
\centering
\includegraphics[width=7cm]{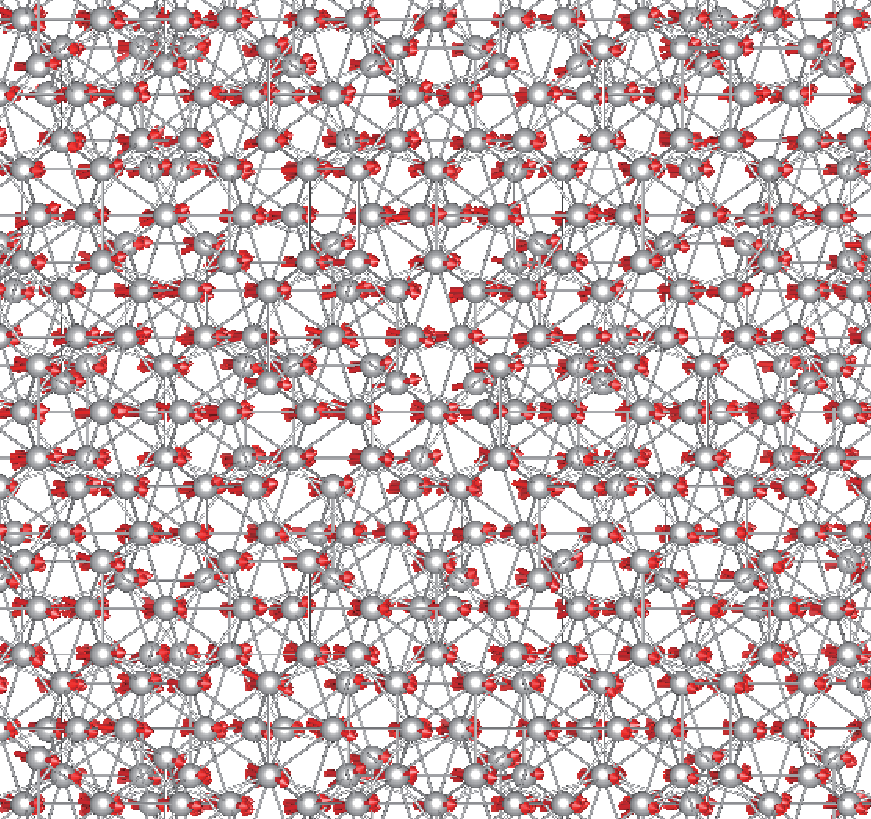}
\caption{
The spin state at $T=0.1$ obtained as a snap shot of the Monte Carlo simulation for the $N=20364$ system. Red arrow denotes the spin at each site illustrated by gray circle. The gray line denotes the bond between the sites with the bond length $5.9821~{\rm \AA}$. 
}
\label{fig:T01_spin}
\end{figure}

We calculated the internal energy $E$ of $H$ in Eq.~(\ref{eq:H}), where the average $\langle E\rangle$ is taken for the inner sites with $N_{\rm IN}$ listed in Table~\ref{tb:SF}. 
The results per site are plotted in Fig.~\ref{fig:E_T}. The error bars are within the symbol sizes at each temperature. 
As temperature decreases, the internal energy decreases. Around $T=T_{\rm c}$, the slope becomes steeper. As $T\to 0$, $\langle E\rangle$ approaches $N_{\rm 5f~bond}J_{\rm 5f}+N_{\rm 2f~bond}J_{\rm 2f}$, where $N_{\rm 5f~bond}$ and $N_{\rm 2f~bond}$ are the numbers of the N.N. bonds and the N.N.N. bonds, respectively. This value is the ground-state energy of the collinear FM state, which will be also confirmed by the snap shot of spin configuration at low temperature obtained by the Monte Carlo simulation (see Fig.~\ref{fig:T01_spin}).  

\begin{figure}[bt]
\centering
\includegraphics[width=7cm]{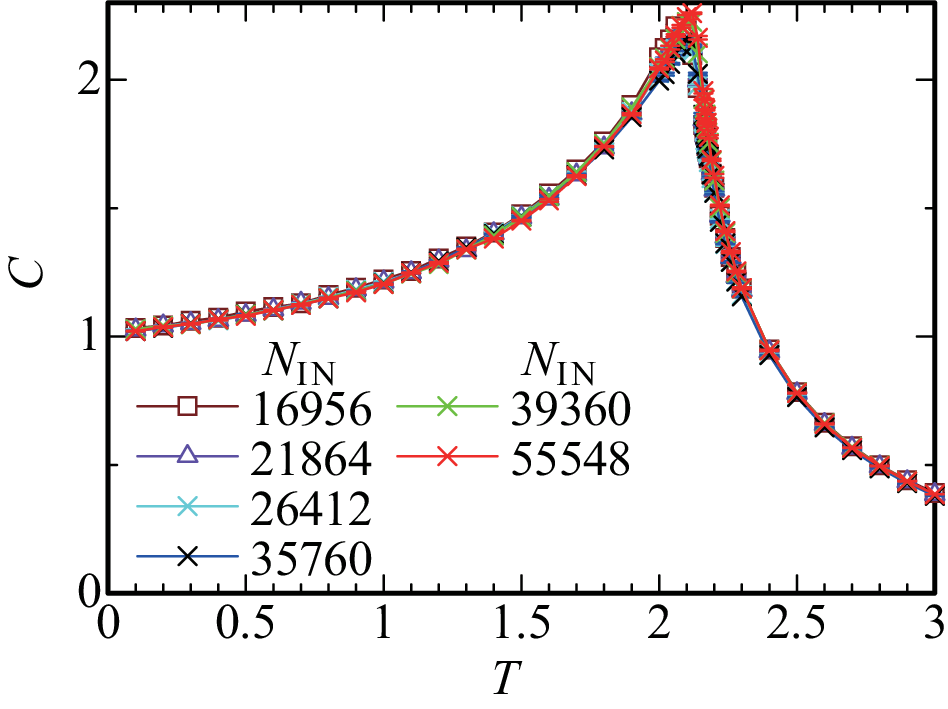}
\caption{
Temperature dependence of specific heat calculated for $J_{\rm 5f}=1$ and $J_{\rm 2f}=1$ in the $N=20364$, 26412, 30048, 39360, 47520 and 62868 systems.
}
\label{fig:C_T}
\end{figure}

Next, we calculated the specific heat 
\begin{eqnarray}
C=\frac{\langle E^2\rangle-\langle E\rangle^2}{N_{\rm IN}T^2},
\label{eq:C}
\end{eqnarray}
which expresses fluctuations of the internal energy.
It is noted that we confirmed that $C(T)$ calculated by Eq.~(\ref{eq:C}) coincides with the temperature derivative of the internal energy $d\langle E\rangle/dT$ [see Fig.~\ref{fig:mcs}(b) in Appendix~\ref{sec:MC_check}].  
The results of the temperature dependences of $C$ for various system sizes are shown in Fig.~\ref{fig:C_T}.
A remarkable peak appears at $T=T_{\rm c}(N_{\rm IN})$, where $T_{\rm c}(N_{\rm IN})$ is defined as the temperature of the location of the peak of $C(T)$ in the system with the inner $N_{\rm IN}$ site, which is reflected in the steep slope in the $\langle E\rangle$-$T$ plot presented in Fig.~\ref{fig:E_T}.

The system size dependences of $C(T)$ in Fig.~\ref{fig:C_T} indicate that the specific heat at $T_{\rm c}$ does not diverge but exhibits a cusp in the bulk limit $N_{\rm IN}\to \infty$. This will be confirmed later by analyzing the critical exponent $\alpha$ of the specific heat 
\begin{eqnarray}
C\sim |T-T_{\rm c}|^{-\alpha}
\label{eq:C_T_alpha}
\end{eqnarray}
in Sec.~\ref{sec:alpha}. Here Eq.~(\ref{eq:C_T_alpha}) expresses the singular part of the specific heat in the vicinity of $T_{\rm c}$. 

The continuous transition is defined as the phase transition where the $n$-th derivative of the free energy for $n\ge 2$ exhibits discontinuity or divergence at $T_{\rm c}$. Since the specific heat is the quantity obtained by the 2nd derivative of the free energy, the result in Fig.~\ref{fig:C_T} indicates that the phase transition in the present system is continuous transition.

\begin{figure}[bt]
\centering
\includegraphics[width=7cm]{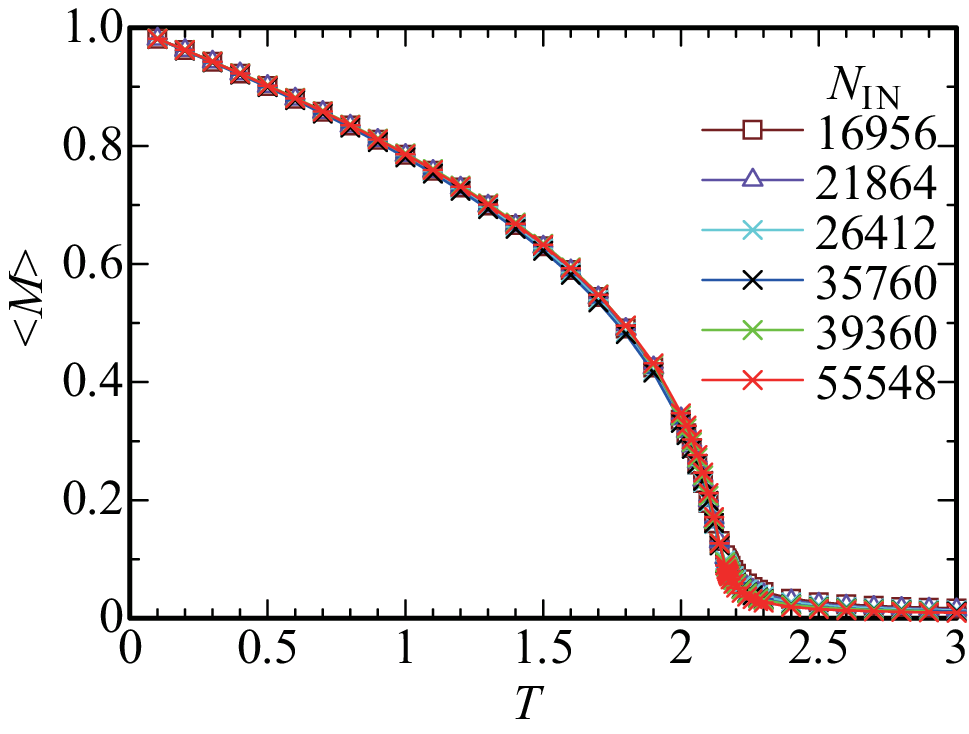}
\caption{
Temperature dependence of magnetization calculated for $J_{\rm 5f}=1$ and $J_{\rm 2f}=1$ in the $N=20364$, 26412, 30048, 39360, 47520 and 62868 systems.
}
\label{fig:M_T}
\end{figure}

Next, we calculated the average of the magnetization $\langle M\rangle$, where $M$ is defined in Eq.~(\ref{eq:M}) and $N$ in Eq.~(\ref{eq:M_mu}) is replaced with $N_{\rm IN}$. 
The temperature dependences of $\langle M\rangle$ in various system sizes are shown in Fig.~\ref{fig:M_T}. 
Around $T=T_{\rm c}$, the magnetization increases with the convex curve and reaches the maximum value for the low-$T$ limit. Namely, the magnetization is fully polarized, i.e., $\langle M\rangle\to 1$ for $T\to 0$, where the collinear FM state is realized in the ground state. In the bulk limit $N_{\rm IN}\to\infty$, the magnetization becomes finite below $T_{\rm c}$, i.e., $\langle M\rangle=0$ for $T>T_{\rm c}$ and $\langle M\rangle\ne 0$ for $T<T_{\rm c}$. Hence, the magnetization is the order parameter of the FM transition. 

The magnetization in the vicinity of $T_{\rm c}$ is expressed as
\begin{eqnarray}
\langle M\rangle\sim(T_{\rm c}-T)^{\beta},
\label{eq:M_T_beta}
\end{eqnarray}
where $\beta$ is the critical exponent. We will estimate $\beta$ by the finite-size scaling in Sec.~\ref{sec:beta}. 

%

The magnetic susceptibility 
\begin{eqnarray}
\chi=\frac{1}{T}\left(\langle{\bm M}\cdot{\bm M}\rangle-\langle{\bm M}\rangle\cdot\langle{\bm M}\rangle\right)
\label{eq:chi_T} 
\end{eqnarray}
expresses fluctuations of the magnetization.  
The singular part of $\chi$ in the vicinity of $T_{\rm c}$ is expressed as
\begin{eqnarray}
\chi\sim |T-T_{\rm c}|^{-\gamma},
\label{eq:chi_gamma}
\end{eqnarray}
where $\gamma$ is the critical exponent. 
Careful analysis of $\gamma$ in the finite-size scaling was discussed for the Heisenberg model on periodic lattices~\cite{Paauw,Holm,Peczak,Chen}. In this study, following the argument in Refs.~\cite{Holm,Peczak,Chen}, we will analyze the criticality of the magnetic susceptibility in the i-QC as below. 
In the high-temperature phase for $T>T_{\rm c}$, the true magnetization vanishes, $\langle{\bm M}\rangle={\bm 0}$. Hence, the magnetic susceptibility where $\langle{\bm M}\rangle$ is set to be ${\bm 0}$ in Eq.~(\ref{eq:chi_T})
\begin{eqnarray}
\chi=\frac{\langle{\bm M}^2\rangle}{T} 
\label{eq:chi_T_M0} 
\end{eqnarray}
can be used for the evaluation of $\gamma$ because as the system size increases, Eq.~(\ref{eq:chi_T_M0}) correctly approaches the thermodynamic limit~\cite{Binder_text}. The detail of the analysis by the finite-size scaling will be given in Sec.~\ref{sec:gamma}.

\begin{figure}[bt]
\centering
\includegraphics[width=7cm]{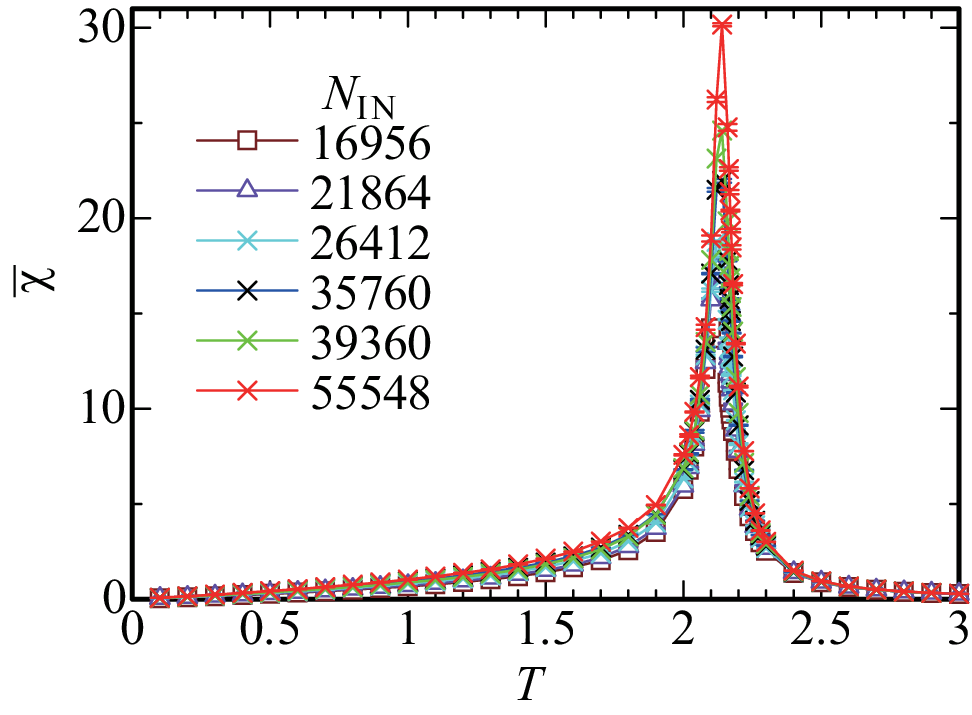}
\caption{
Temperature dependence of magnetic susceptibility $\bar{\chi}$ calculated for $J_{\rm 5f}=1$ and $J_{\rm 2f}=1$ in the $N=20364$, 26412, 30048, 39360, 47520 and 62868 systems.
}
\label{fig:chi_abs_T}
\end{figure}

To extract the enhancement of the magnetic susceptibility at $T_{\rm c}(N_{\rm IN})$ in finite-size systems, following Refs.~\cite{Holm,Peczak,Chen,Solanki},   
we define the magnetic susceptibility 
\begin{eqnarray}
\bar{\chi}=\frac{1}{T}(\langle{\bm M}^2\rangle-\langle M\rangle^2),
\label{eq:chi_abs}
\end{eqnarray}
where the second term of the numerator is the average of the absolute value of the magnetization vector $M=|{\bm M}|$. 
The temperature dependences of $\bar{\chi}$ in various system sizes are shown in Fig.~\ref{fig:chi_abs_T}. 
At $T=T_{\rm c}(N_{\rm IN})$, a sharp peak appears and the hight of the peak develops as the system size increases. 
For $T\to 0$, the magnetization saturates $\langle M\rangle\to 1$ as shown in Fig.~\ref{fig:M_T} and the average of the square of the magnetization vector also leads to $\langle {\bm M}^2\rangle \to 1$. Hence, as $T$ decreases toward absolute zero, $\bar{\chi}(T)$ is suppressed as $\bar{\chi}(T)\to 0$. 

To see the spin alignment in real space, here we show the snapshot of the spin state obtained by the Monte Carlo simulation. 
Figure~\ref{fig:T01_spin} shows the spin state at $T=0.1$ in the $N=20364$ system. The spins are denoted by red arrows on the sites. Here we show the same lattice geometry presented in Fig.~\ref{fig:QC_8_class} which is the view from the 5-fold axis direction. We see that all spins tend to be aligned collinearly toward the right direction.

The results obtained in the finite system sizes shown in Fig.~\ref{fig:E_T}, Fig.~\ref{fig:C_T}, Fig.~\ref{fig:M_T}, and Fig.~\ref{fig:chi_abs_T} indicate that the phase transition to the collinear FM order occurs in the bulk limit, which is validated by the analysis of the Binder parameter concluding $T_{\rm c}=2.1725$, as described in Sec.~\ref{sec:Binder}. In the following subsections, we will perform the finite-size scaling analysis to obtain the critical exponents. 

\subsection{Critical exponent of spin correlation length $\nu$}
\label{sec:spin_corr}

Near the transition temperature, the spin correlation length $\xi$ diverges as~\cite{Barber}
\begin{eqnarray}
\xi\sim |T-T_{\rm c}|^{-\nu}, 
\label{eq:def_nu}
\end{eqnarray}
where $\nu$ is the critical exponent of spin correlation lengths.

\begin{figure}[bt]
\centering
\includegraphics[width=7cm]{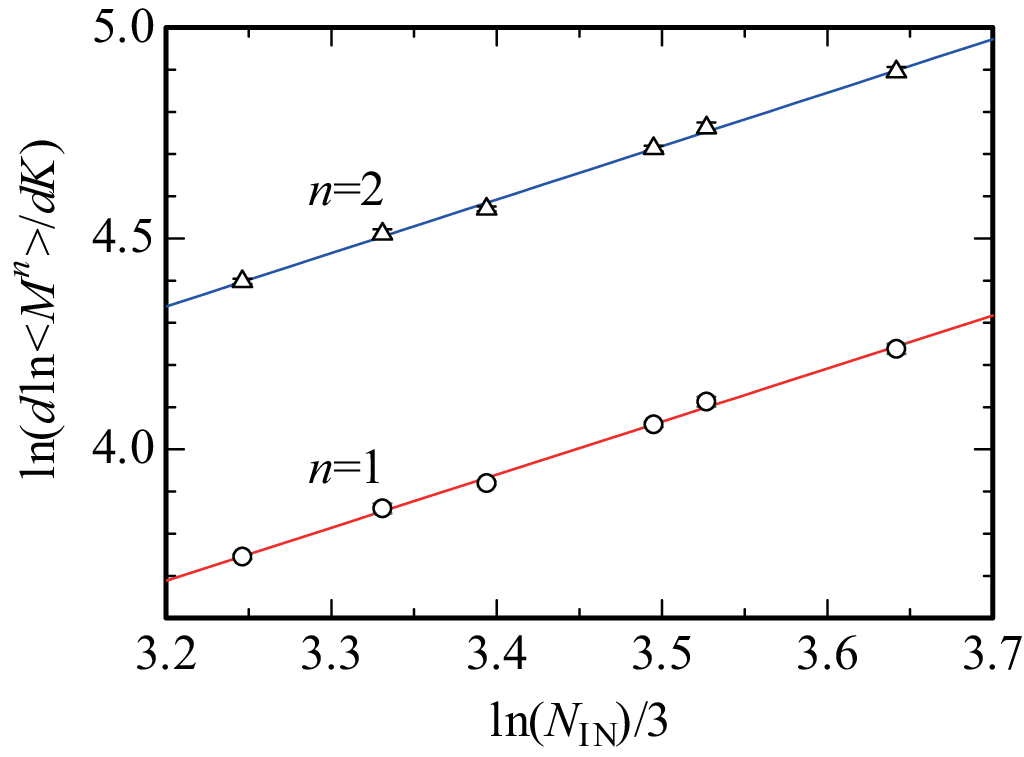}
\caption{
Thermodynamic derivative $\frac{d\ln\langle M^n\rangle}{dK}$ calculated at $T_{\rm c}=2.1725$ vs lattice size $L$ in a double-logarithmic plot for $n=1$ (circle) and $n=2$ (triangle). 
}
\label{fig:dlnMdK}
\end{figure}

It is known~\cite{Ferrenberg,Solanki} that the derivative of cumulant of the magnetization at $T_{\rm c}$ obeys 
\begin{eqnarray}
\frac{d\ln\langle M^n\rangle}{dK}\sim L^{1/\nu}, 
\label{eq:dlnMdK}
\end{eqnarray}
where $n$ is an integer and the inverse temperature $K$ is defined by $K\equiv 1/T$. 
In general, temperature derivative of the average of the physical quantity can be expressed as the average of correlation of the energy and the physical quantity. 
The left hand side of Eq.~(\ref{eq:dlnMdK}) is calculated as
\begin{eqnarray}
\frac{d\ln\langle M^n\rangle}{dK}=\langle E\rangle-\frac{\langle EM^n\rangle}{\langle M^n\rangle}.  
\label{eq:EM}
\end{eqnarray}
In order to estimate $\nu$, we calculate the right hand side of Eq.~(\ref{eq:EM}) for $n=1$ and 2 in various system sizes. The results for $n=1$ and 2 at $T_{\rm c}$ are shown by the open circles and open triangles respectively in the ln-ln plot in Fig.~\ref{fig:dlnMdK}. 
We set the linear size $L$ as $N_{\rm IN}^{1/3}$, i.e., $L=N_{\rm IN}^{1/3}$. 
The linear fit of the data for $n=1$ gives the inverse slope as  
%
$1/\nu=1.2580\pm0.0320$.
%
The linear fit of the data for $n=2$ gives the inverse slope as
%
$1/\nu=1.2674\pm 0.0290$.
%
Following the argument in Ref.~\cite{Peczak}, from these two estimates we obtain 
\begin{eqnarray}
\nu=0.792\pm 0.017.
\label{eq:nu}
\end{eqnarray}
%

\subsection{Critical exponent of the magnetization $\beta$}
\label{sec:beta}

\begin{figure}[bt]
\centering
\includegraphics[width=7cm]{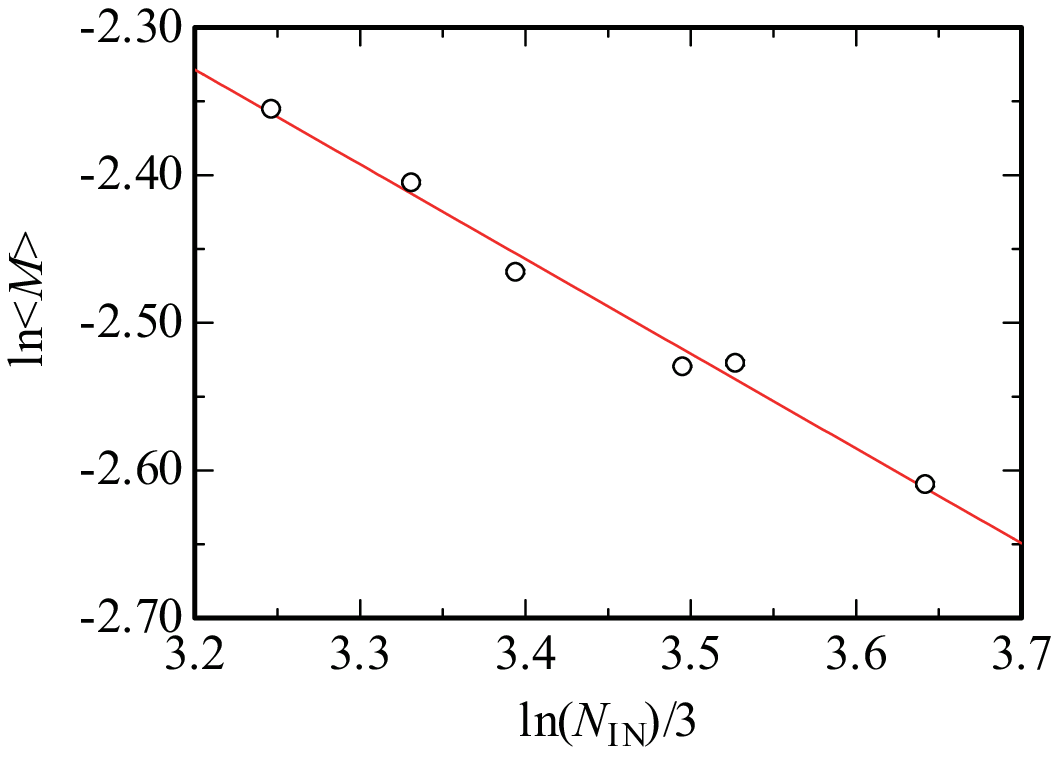}
\caption{
Magnetization $\langle M\rangle$ calculated at $T_{\rm c}=2.1725$ vs lattice size $L$ in a double-logarithmic plot.
}
\label{fig:lnM_lnL}
\end{figure}

The magnetization at the transition temperature $T=T_{\rm c}$ scales as~\cite{Barber} 
\begin{eqnarray}
\langle M\rangle\sim L^{-\beta/\nu}
\label{eq:M_bn}
\end{eqnarray}
where $\beta$ is the critical exponent of the magnetization. 
To extract the ratio of the critical exponents $\beta/\nu$, 
we plot $\ln\langle M\rangle$ vs $\ln{L}$ in Fig.\ref{fig:lnM_lnL}, where we set $L=N_{\rm IN}^{1/3}$.  The slope of the straight line is estimated by the linear least-square fit as 
\begin{eqnarray}
\beta/\nu=0.64171\pm0.0347.
\label{eq:bn}
\end{eqnarray}
By using the value of $\nu$ estimated as Eq.~(\ref{eq:nu}) in Eq.~(\ref{eq:bn}), we obtain
%
$\beta=0.508(30)$.
%

\subsection{Critical exponent of the magnetic susceptibility $\gamma$}
\label{sec:gamma}

\begin{figure}[bt]
\centering
\includegraphics[width=7cm]{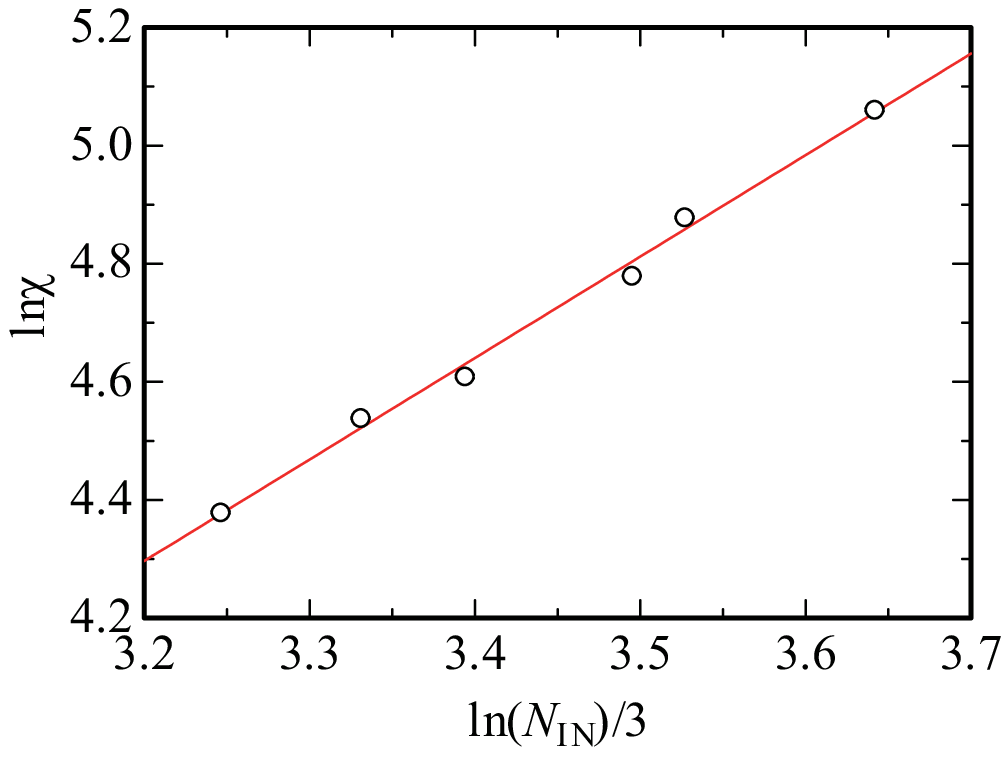}
\caption{
Magnetic susceptibility $\chi$ calculated at $T_{\rm c}=2.1725$ vs lattice size $L$ in a double-logarithmic plot. 
}
\label{fig:ln_chi_lnL}
\end{figure}

The magnetic susceptibility at the transition temperature $T=T_{\rm c}$ scales as~\cite{Barber} 
\begin{eqnarray}
\chi\sim L^{\gamma/\nu}, 
\label{eq:chi_gn}
\end{eqnarray}
where $\gamma$ is the critical exponent of the magnetic susceptibility. To extract the ratio of the critical exponents $\gamma/\nu$, 
we plot $\ln\chi$ vs $\ln N_{\rm IN}^{1/3}$ in Fig.~\ref{fig:ln_chi_lnL}.  Here we used $\chi$ defined in Eq.~(\ref{eq:chi_T_M0}). The slope of the straight line is estimated by the linear least-square fit as 
%
\begin{eqnarray}
\gamma/\nu=1.7183\pm0.0648. 
\label{eq:gn}
\end{eqnarray}
By using the value of $\nu$ estimated as Eq.~(\ref{eq:nu}) in Eq.~(\ref{eq:gn}), we obtain
%
$\gamma=1.361(59)$.
%

From Eq.~(\ref{eq:bn}) and Eq.~(\ref{eq:gn}), we obtain $2\beta/\nu+\gamma/\nu=3.0017\pm 0.1342$. When the hyper scaling relation holds, the critical exponents and the spatial dimension $d$ should satisfy the relation $2\beta/\nu+\gamma/\nu=d$~\cite{Solanki}, where $d=3$ in the present system. Hence, our results satisfy the hyper scaling relation 
within the margin of error. 

With use of the scaling relation $\eta=2-\gamma/\nu$, from the result of Eq.~(\ref{eq:gn}) we obtain the critical exponent  
$\eta=0.2817\pm 0.0648$. 

We also estimate the other critical exponent $\delta$ with use of the scaling relation $\delta=\frac{\beta+\gamma}{\beta}$. By inputting the values of $\beta/\nu$ in Eq.~(\ref{eq:bn}) and $\gamma/\nu$ in Eq.~(\ref{eq:gn}) into this equation, we obtain $\delta=\frac{\beta/\nu+\gamma/\nu}{\beta/\nu}=3.68(23)$.  

\subsection{Critical exponent of the specific heat $\alpha$}
\label{sec:alpha}

The specific heat at the transition temperature $T=T_{\rm c}$ scales as
\begin{eqnarray}
C\sim C_{\rm reg}+aL^{\alpha} 
\label{eq:C_a}
\end{eqnarray}
with the critical exponent $\alpha$, where $C_{\rm reg}$ is a regular background term and $a$ is a coefficient. 
In the analyses of the critical exponents $\beta$ and $\gamma$, direct evaluations of $\langle M\rangle$ and $\chi$ as well as $\bar{\chi}$ made it possible to extract the values of $\beta/\nu$ and $\gamma/\nu$ in the ln-ln plots of Eq.~(\ref{eq:M_bn}) and Eq.~(\ref{eq:chi_gn}) respectively. 
However, in the case of the specific heat, large contribution from the non-universal background terms $C_{\rm reg}$ makes it difficult to extract the value of $\alpha$. 

Hence, we evaluate $\alpha$ by using the hyper scaling relation $\alpha=2-d\nu$. By employing the value of $\nu$ estimated in Eq.~(\ref{eq:nu}), we obtain $\alpha=-0.376(51)$. 
The negative value of $\alpha$ indicates that the specific heat does not diverge at $T=T_{\rm c}$ but has a cusp (see Fig.~\ref{fig:C_T}). This implies that the sign of the coefficient $a$ in Eq.~(\ref{eq:C_a}) is negative. 

It is noted that our estimates of $\alpha=-0.376(51)$, $\beta=0.508(30)$, and $\gamma=1.361(59)$ yield $\alpha+2\beta+\gamma=2.001\pm0.158$, which satisfies the Rushbrook's relation~\cite{Rushbrooke} $\alpha+2\beta+\gamma=2$ within the margin of error. 

\subsection{Universality class of Heisenberg model on icosahedral quasicrystal}
\label{sec:UC}

The critical exponents obtained in this study is summarized in Table~\ref{tb:CE}. 
The critical exponents in the classical Heisenberg model on the 3D periodic crystal~\cite{Stanley} and spin-glass system~\cite{Ogawa} are also listed in Table~\ref{tb:CE} for comparison. The critical exponent $\gamma$ in the i-QC is the similar value of the periodic crystals while $\nu$ is larger than the value of the periodic crystal but is much smaller than that in the 3D spin glass system. 
The critical exponent $\beta$ is close to the mean-field value. The critical exponent $\eta$ is estimated to be one-order of magnitude larger than that in periodic crystals.

\onecolumngrid

\begin{table}[b]
\begin{tabular}{ccccccc}
\hline
class & $\alpha$ & $\beta$ & $\gamma$ & $\delta$ & $\nu$ & $\eta$ \\
\hline
3D Heisenberg & $-0.12$ & $0.36$ & $1.39$ & $4.9$ & $0.71$ &  $0.04$ \\
3D spin glass & $-1.6$ & $0.45$ & $2.7$ & $7.0$ & $1.2$ & $-0.25$ \\
icosahedral quasicrystal & $-0.376(51)$ & $0.508(30)$ & $1.361(59)$ & $3.68(23)$ & $0.792(17)$ & $0.282(65)$\\
\hline
\end{tabular}
\caption{Critical exponents for universality class of the 3D Heisenberg model in periodic crystal~\cite{CL_text}, spin glass~\cite{Ogawa}, and icosahedral quasicrystal.}
\label{tb:CE}%
\end{table}

\twocolumngrid

\subsection{Magnetization and correlation length for each site class}
\label{sec:M_LE}

\begin{figure}[bt]
\centering
\includegraphics[width=7cm]{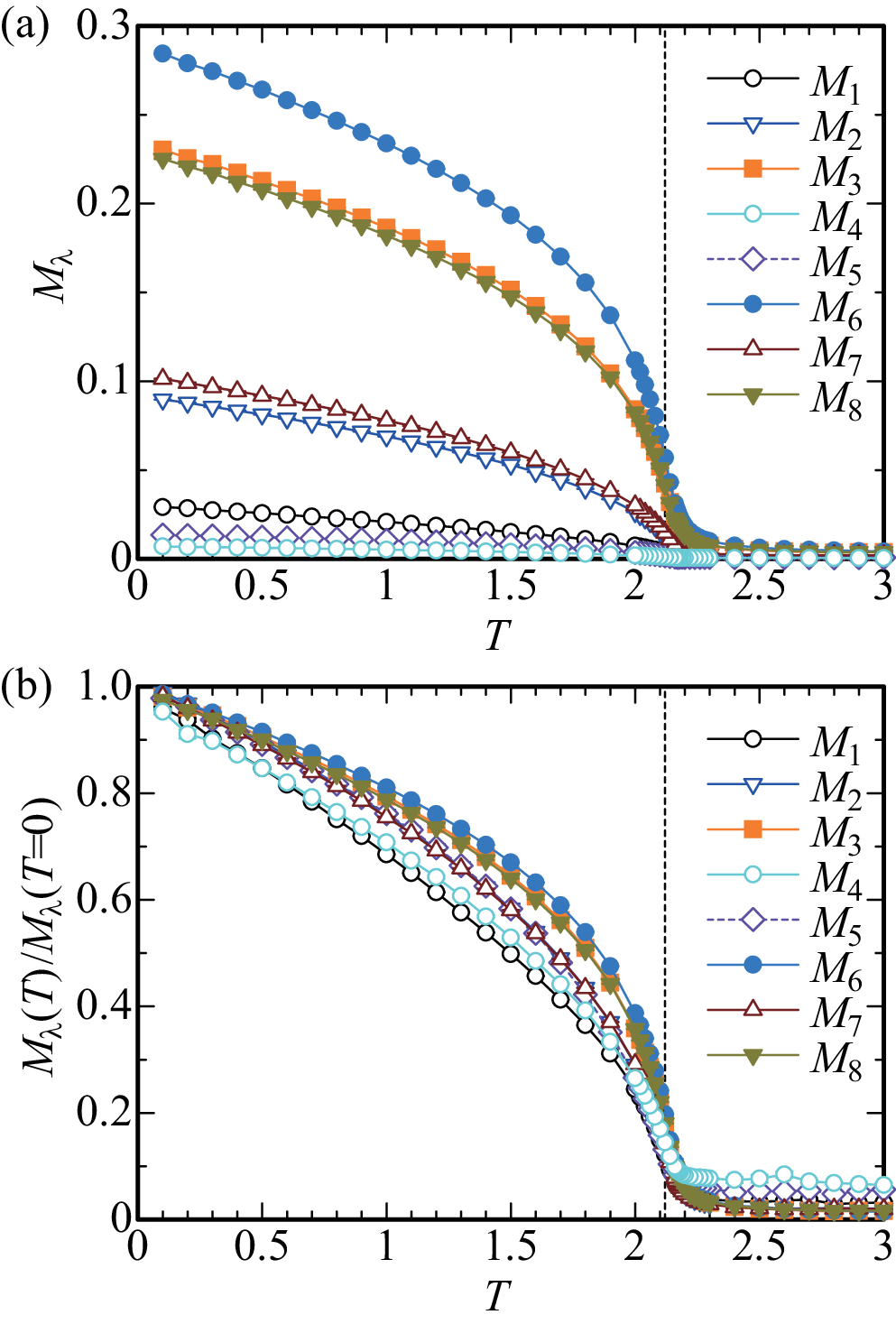}
\caption{(a) Temperature dependence of site-dependent magnetization $M_{\lambda}$ where $\lambda$ is the label classifying each site by local environment in the $N=62868$ system. (b) Temperature dependence of $M_{\lambda}$ normalized by each frequency $N_{\lambda}/N_{\rm IN}=M_{\lambda}(T\to 0)$ which is equal to the lowest-temperature limit of magnetization for each class. In (a) and (b), a vertical dashed line denotes $T_{\rm c}(N_{\rm IN})$. 
}
\label{fig:M_LE}
\end{figure}

To gain further insight into the magnetism of the i-QC, we calculate the magnetization for each site class (see Table~\ref{tb:LE} and Fig.~\ref{fig:QC_8_class}) 
$M_{\lambda}$ defined by 
\begin{eqnarray}
{\bm M}_{\lambda}=\frac{1}{N_{\rm IN}}\sum_{i=1}^{N_{\lambda}}{\bm S}_i, 
\label{eq:M_LE}
\end{eqnarray}
where $\lambda$ labels each class with $N_{\rm IN}=\sum_{\lambda=1}^{8}N_{\lambda}$. The average of the Monte Carlo sampling is taken for the inner sites.  The result of the temperature dependence of $M_{\lambda}$ for the $N=62868$ system is shown in Fig.~\ref{fig:M_LE}(a). As temperature decreases, the magnetization of each class $M_{\lambda}(T)$ increases simultaneously below $T_{\rm c}(N_{\rm IN})$ with $N_{\rm IN}=55548$ indicated by the vertical dashed line. 
This is more clearly visible when we plot each magnetization scaled by each frequency $N_{\lambda}/N$ in Fig.~\ref{fig:M_LE}(b). All the scaled magnetization for $\lambda=1,\cdots,8$ start to increase sharply below $T_{\rm c}(N_{\rm IN})$, indicating cooperative development of each magnetization $M_{\lambda}$. 
For the low-temperature limit, the magnetization for each class reaches the maximum as $\lim_{T\to 0}M_{\lambda}(T)=N_{\lambda}/N_{\rm IN}$, indicating full polarization at each site. 
It is noted that convex curves are drawn with the filled symbols ($\lambda=3, 6,$ and $8$) being almost overlapped for whole temperatures, whose densities share $76\%$ in total. 
The convex curves are also drawn with the open symbols ($\lambda\ne 3, 6,$ and $8$), whose trajectories are beneath those with the filled symbols.  

\begin{figure}[bt]
\centering
\includegraphics[width=7cm]{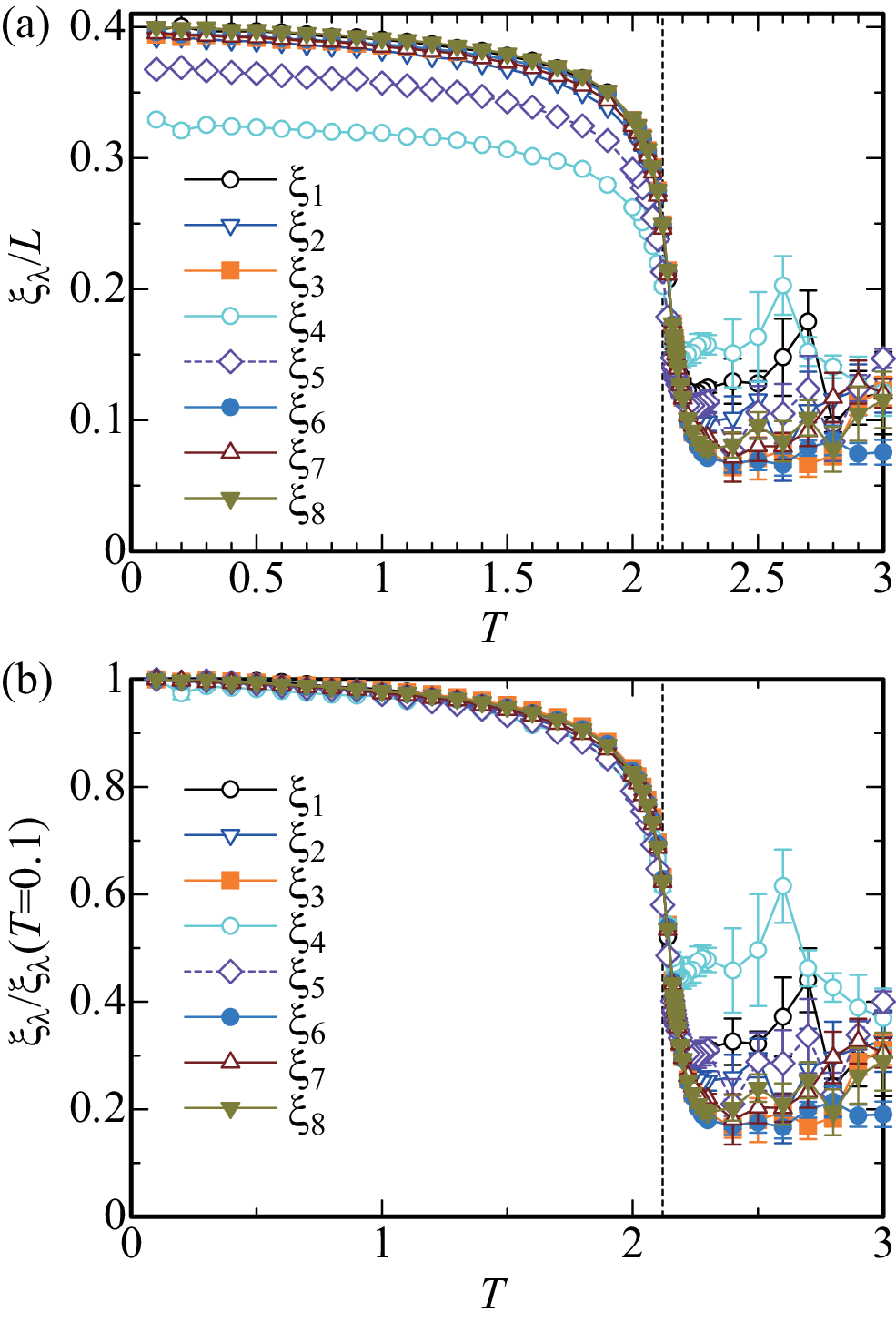}
\caption{
(a) Temperature dependence of spin correlation length ratio $\xi_{\lambda}/L$ for each class $\lambda$ in the $N=62868$ system. (b) Temperature dependence of spin correlation length normalized by the value at $T=0.1$ for each class.   In (a) and (b), a vertical dashed line denotes $T_{\rm c}(N_{\rm IN})$. 
}
\label{fig:xi_LE}
\end{figure}

We also calculated the temperature dependence of the correlation length $\xi_{\lambda}$ between the spins at the sites belonging to the same class $\lambda$ on the basis of Eq.~(\ref{eq:corrl}). The result of the temperature dependence of $\xi_{\lambda}(T)$ scaled by the diameter $L$ of the system with $N=62868$ is shown in Fig.~\ref{fig:xi_LE}(a). 
At high temperatures than $T_{\rm c}(N_{\rm IN})$, the data scatter with large error bars, while 
below $T_{\rm c}(N_{\rm IN})$ denoted by the vertical dashed line, all correlation length $\xi_{\lambda}$ $(\lambda=1,\cdots, 8)$ increases simultaneously as $T$ decreases. 
In Fig.~\ref{fig:xi_LE}(b), we plot the temperature dependence of the spin correlation length scaled by the one at the lowest temperature $T=0.1$ which is regarded as the correlation length of the sites for the class $\lambda$. As shown in Fig.~\ref{fig:QC_8_class}, each site labeled $\lambda=1,\cdots,8$ is distributed in the intermixed manner. Since for $T\to 0$ all spins are aligned collinearly, $\xi_{\lambda}(T\to 0)$ is regarded as the effective distance of each site with the same class $\lambda$. Hence, the spin correlation length $\xi_{\lambda}(T)$ scaled by the effective distance for each class is interpreted as the mean value of the spin correlation for the site class $\lambda$. 
For $T<T_{\rm c}$, all $\xi_{\lambda}(T)/\xi_{\lambda}(T=0.1)$ follow the same trajectory with steep increase just below $T_{\rm c}(N_{\rm IN})$ indicating the cooperative phenomena.

%

As temperature decreases, spin correlation for each site class evolves and  
when all the spin correlation length normalized by the effective distance of each site class exceeds a threshold, the magnetization scaled by the frequency for all site class $\lambda$ starts to increase simultaneously. 
These results imply that the difference in local environment of each site is governed by cooperative development of spin correlations on cooling, which gives rise to the FM transition. 
%

\section{Summary and discussion}
In summary, we have studied the nature of the FM phase transition in the i-QC by performing the Monte Carlo simulation for the classical Heisenberg model in the Cd$_{5.7}$ Yb. 
By applying the finite-size scaling to the Heisenberg model for the N.N. interaction $J_{\rm 5f}=1$ and the N.N.N. interaction $J_{\rm 2f}=1$, we have identified the FM transition temperature $T_{\rm c}=2.1725$ and the critical exponents $\nu=0.792(17)$, $\beta=0.508(30)$, and $\gamma=1.361(59)$. Within the statistical error, we have confirmed that the hyper scaling relation holds. With use of the hyper scaling relation, we have obtained $\alpha=-0.376(51)$, $\delta=3.68(23)$, and $\eta=0.282(65)$. The obtained results of $\alpha$, $\beta$, and $\gamma$ satisfy the Rushbrooke's scaling relation within the statistical error.  
From these results, we have revealed the universality class of the i-QC, which is ditinct from those of the periodic crystal and the spin glass.  

In this study, to clarify the universality class of the i-QC, we studied the Heisenberg model for $J_{\rm 5f}=1$ and $J_{\rm 2f}=1$ as a typical parameter. As noted in Fig.~\ref{fig:IC_QC}(c), the bond lengths of the 5-fold-axis and 2-fold-axis directions are about $5\%$ different. Hence, the $J_{\rm 2f}/J_{\rm 5f}$ ratio may be slightly deviate from 1 in real materials. However, since the variation of $J_{\rm 2f}/J_{\rm 5f}$ is small, the critical exponents clarified in this paper, i.e., the list for the i-QC in Table~\ref{tb:CE}, are considered to capture the essential values. 

In the Cd$_{5.7}$Yb, each Yb site at the vertex of the icosahedron is classified into 8 class with respect to the coordination numbers of the N.N. and N.N.N. bonds. As temperature decreases, spin correlation for each site class evolves individually. When all the correlation length scaled by the effective distance for each site class exceed the threshold, all the magnetization starts to increase, which gives rise to the FM transition.  

Recently, experimental identification of the critical exponents has been reported in the Gd-based approximant crystals~\cite{Shiino2022}. 
In Au$_{72.7}$Si$_{13.6}$Gd$_{13.7}$,  $\beta=0.47$, $\gamma=1.12$, and $\delta=3.60$ were identified and in Au$_{68.6}$Si$_{16.0}$Gd$_{15.4}$,  $\beta=0.51$, $\gamma=1.00$, and $\delta=3.38$ were identified~\cite{Shiino2022}. In both materials, the critical exponents of the magnetization $\beta$ were close to the mean field value $\beta=0.5$. 
The present study based on the i-QC composed of regular icosahedrons has explained the value of $\beta=0.5$ observed in the 1/1 approximant crystals which consist of icosahedrons. 

The FM long-range order was discovered in the i-QC Au-Ga-Gd~\cite{Tamura2021}. It is interesting to observe the critical exponents experimentally to make a comparison with our theoretical results.  The Gd-based QC with Gd$^{3+}$ and also Eu-based QC with Eu$^{2+}$ have $4f^7$ configuration whose ground multiplet is $^8S_{7/2}$. These materials are candidates for comparison with our theoretical results, where the magnetic anisotropy arising from the CEF is absent.  



\begin{acknowledgments}
The author (S.W.) is deeply indebted to R. Eto and S. Suzuki for useful discussion about implementation of the Monte Carlo method.
The authors acknowledge K. Momma for expanding functionality of VESTA to draw the magnetic moments as well as atoms in the QC~\cite{Momma}. 
This work was supported by JSPS KAKENHI Grant Numbers JP22H01170, JP23K17672, JP24H01675, JP19H05819, JP19H05818, and JP24K08041.
\end{acknowledgments}



\appendix

\section{Monte Carlo steps}
\label{sec:MC_check}

\begin{figure}[h]
\centering
\includegraphics[width=7cm]{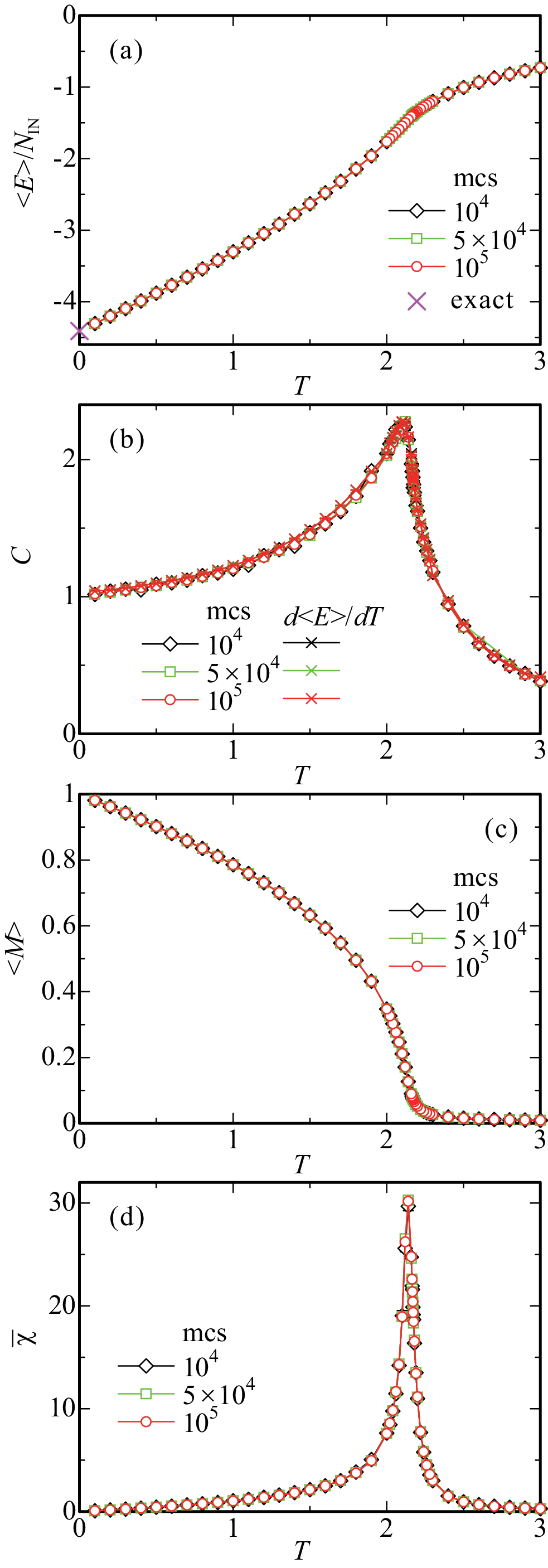}
\caption{
Temperature dependences of (a) internal energy per site $\langle E\rangle/N_{\rm IN}$, (b) specific heat $C$, (c) magnetization $\langle M\rangle$, and (d) magnetic susceptibility $\bar{\chi}$ for $J_{\rm 5f}=1$ and $J_{\rm 2f}=1$ obtained by the Monte Carlo simulations with the mcs $10^4$ (diamond), $5\times 10^4$ (square), and $10^5$ (circle) in the $N=62868$ system. In (a), the exact ground-state energy for the collinear FM state $\epsilon_{0}$ is denoted by cross symbol. In (b), $d\langle E\rangle/dT$ calculated with mcs $10^4$ (black cross), $5\times 10^4$ (green cross), and $10^5$ (red cross) are plotted. 
}
\label{fig:mcs}
\end{figure}

In this appendix, we show the Monte Carlo steps (mcs) dependence of physical quantities in our Monte Carlo simulations. 
As noted in Sec.~\ref{sec:MC}, we performed the Monte Carlo simulation based on the heat-bath method and the over-relaxation method, which are combined with the replica exchange method of parallel tempeeing. 
We calculated the internal energy $\langle E\rangle/N_{\rm IN}$, the specific heat $C$, the magnetization $M$, and the magnetic susceptibility $\bar{\chi}$ for $J_{\rm 5f}=1$ and $J_{\rm 2f}=1$ with the mcs $10^4$ (diamond), $5\times 10^4$ (square), and $10^5$ (circle) in the $N=62868$ system, as shown in Figs.~\ref{fig:mcs}(a), \ref{fig:mcs}(b), \ref{fig:mcs}(c), and \ref{fig:mcs}(d), respectively. Here, the average of the physical quantities are taken for the inner sites with $N_{\rm IN}=55548$. The error bars are within the size of each symbol at each temperature in Figs.~\ref{fig:mcs}(a)-(d).

In Fig.~\ref{fig:mcs}(a), we see that all the symbols are overlapped each other at each temperature. The lowest temperature limit of the data reaches the exact value of the ground-state energy per site 
\begin{eqnarray}
\epsilon_0=\frac{N_{\rm 5f \ bond}J_{\rm 5f}+N_{\rm 2f \ bond}J_{\rm 2f}}{N_{\rm IN}} 
\label{eq:E_exact}
\end{eqnarray}
of the collinear FM state (cross), where $N_{\rm 5f \ bond}$ is the number of the N.N. bonds and $N_{\rm 2f \ bond}$ is the number of the N.N.N. bonds. These results imply that the calculation with the mcs $10^4$ already provide the correct value of the internal energy. 

In Fig.~\ref{fig:mcs}(b), we plot the temperature dependence of the specific heat $C$. We also plot the temperature derivative of the internal energy $\frac{d\langle E\rangle}{dT}$ (cross) by performing a finite-difference derivative of the $\langle E\rangle$ data. We see the similar curves of $C$ and $\frac{d\langle E\rangle}{dT}$, although the latter is obtained by a finite-difference derivative numerically, which confirms that the calculation of the specific heat by Eq.~(\ref{eq:C}), i.e., fluctuations of the energy is correctly done. The data of $C$ for mcs $10^4$, $5\times 10^4$, and $10^5$ are overlapped in the whole temperature range. 

In Fig.~\ref{fig:mcs}(c), we see that all symbols at each temperature are overlapped, which indicates that the mcs $10^4$ is enough to calculate the magnetization $\langle M\rangle$ for whole temperatures. 

In Fig.~\ref{fig:mcs}(d), we plot the temperature dependence of $\bar{\chi}$. We see that all results for mcs $10^4$, $5\times 10^4$, and $10^5$ shows the similar curves in the whole temperature range 
and the symbols for mcs $5\times 10^4$ and $10^5$ are overlapped even near $T_{\rm c}(N_{\rm IN})$. 

From these results, we confirmed that the mcs $10^5$ is sufficient for the calculations of the physical quantities. Then we performed the Monte Carlo simulations with the mcs $10^5$ in the present study.

\section{Effects of surfaces on physical quantities}
\label{sec:SF}

In this appendix, we discuss the effect of the surface of the i-QC in the Monte Carlo simulation. As explained in section~\ref{sec:inner}, we performed the Monte Carlo simulation for the system with $N$ sites. To reduce the surface effects, we took the average of the physical quantities for the inner $N_{\rm IN}$ sites listed in Table~\ref{tb:SF}. Here we compare the results obtained by the Monte Carlo sampling for the total sites and for the inner sites.

\begin{figure}[tb]
\centering
\includegraphics[width=7cm]{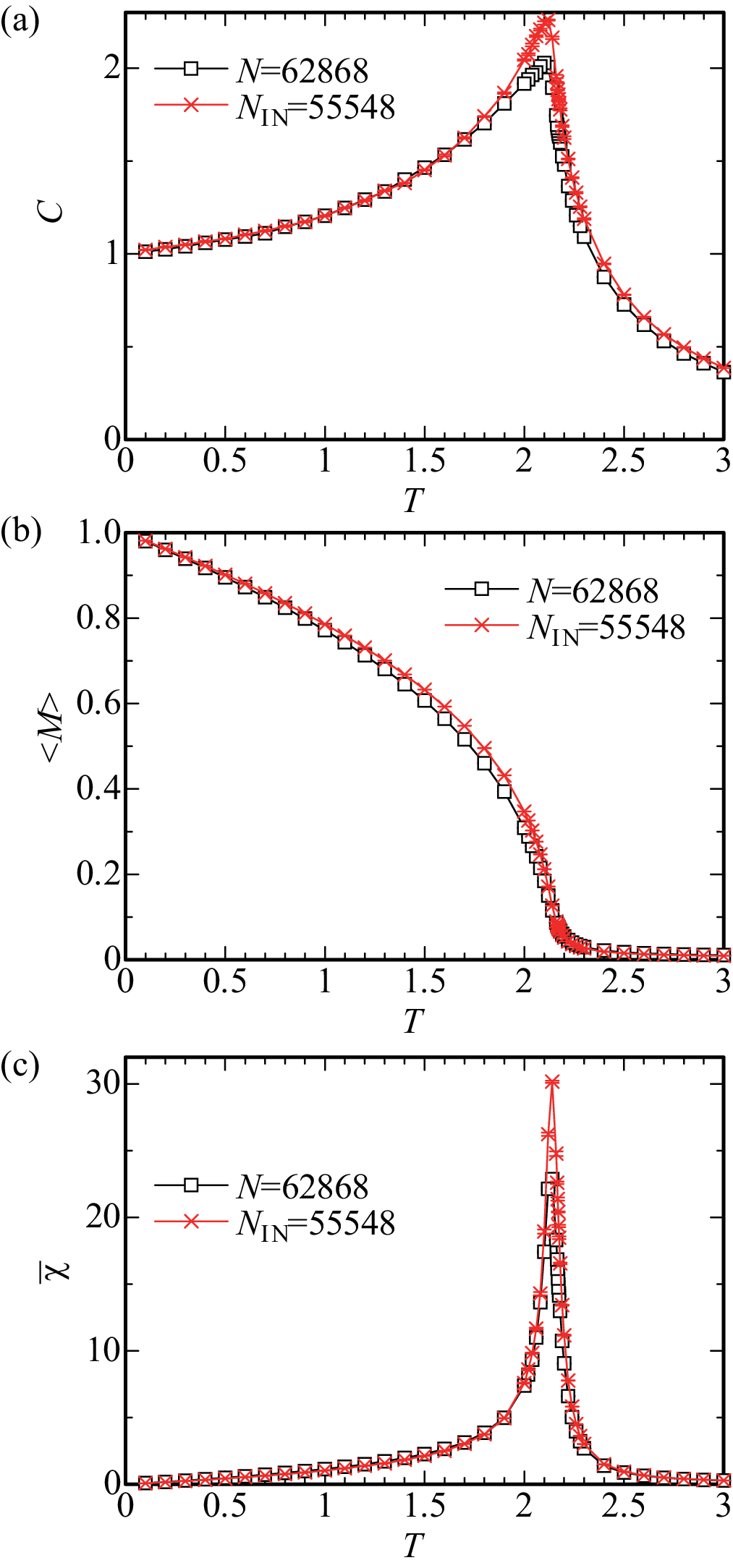}
\caption{
Temperature dependences of (a) specific heat $C$, (b) magnetization $\langle M\rangle$, and magnetic susceptibility $\bar{\chi}$ for $J_{\rm 5f}=1$ and $J_{\rm 2f}=1$ in the $N=62868$ system. Average for the Monte Carlo sampling is taken in the total sites (open square) and in the inner sites with $N_{\rm IN}=55548$ (cross).
}
\label{fig:SF_comp}
\end{figure}

We show both the data of the temperature dependences of the specific heat $C$, the magnetization $\langle M\rangle$, and the magnetic susceptibility $\chi$ in Figs.~\ref{fig:SF_comp}(a), \ref{fig:SF_comp}(b), and \ref{fig:SF_comp}(c), respectively in the $N=62868$ system as a representative case. The data obtained by the Monte Carlo sampling for the total sites $N=62868$ are denoted by open square symbols and for the inner sites $N_{\rm IN}=55548$ are denoted by cross symbols. The error bars are within the sizes of each symbol at each temperature. 

Figure~\ref{fig:SF_comp}(a) shows that the specific heat $C(T)$  around $T_{\rm c}$ calculated for the inner sites is slightly larger than that calculated for the total sites. 
The magnetization $\langle M(T)\rangle$ calculated for the total sites and the inner sites are almost overlapped in the vicinity of $T_{\rm c}$ and $T\to 0$, while a slight deviation appears in the intermediate temperature regime below $T_{\rm c}$, as shown in Fig.~\ref{fig:SF_comp}(b).
 The magnetic susceptibility $\bar{\chi}$ calculated for the inner sites is slightly larger than that calculated for the total sites around $T_{\rm c}$. 

We also made a comparison in the $N=20364$ system with the inner $N_{\rm IN}=16956$ sites. The similar tendency as above was seen in the data of the specific heat $C$ and the magnetization $\langle M\rangle$, while the data of the magnetic susceptibility $\bar{\chi}$ calculated for the total sites and the inner sites are almost overlapped around $T_{\rm c}$.

\end{document}